\documentclass[english, 12pt]{article}
\usepackage[T1]{fontenc}
\usepackage{babel}
\usepackage{ifsym}
\usepackage{geometry}
\usepackage{amsmath, amsthm, amssymb}
\usepackage{graphicx}
\usepackage{booktabs}
\usepackage{rotating}
\usepackage{setspace}
\usepackage{authblk}  
\usepackage{epigraph}
\usepackage{enumitem}   
\usepackage{url}
\usepackage[round]{natbib}
\usepackage{xcolor}

\usepackage[colorlinks,
linkcolor=red,
anchorcolor=blue,
citecolor=blue
]{hyperref}

\newtheorem*{theorem*}{Theorem}

\newtheorem{lemma}{Lemma}

\newtheorem{Definition}{Definition}

\newtheorem{Proposition}{Proposition}

\makeatletter
\usepackage{babel}
\usepackage{bbm}
\usepackage{titlesec}
\titleformat*{\subsubsection}{\normalfont\fontsize{12}{17}\itshape}
\usepackage[toc,page]{appendix}

\makeatother
\begin{document}
	
	\title{Motivating Careerists\thanks{This paper is a revision of the first chapter of my dissertation. I am especially grateful to Peter Buisseret for his encouragement and support during the Covid-19 pandemic. I also thank Scott Ashworth, Ethan Bueno de Mesquita, Justin Fox, Dimitri Landa, Barton Lee, Clement Minaudier, Greg Sheen, Tara Slough, Congyi Zhou, seminar participants of the IBEO workshop, NYU political economy workshop, MPSA annual conference, and APSA annual conference, for their valuable comments. All errors are mine. }
 
 }
	\pagenumbering{gobble}

	\author{Liqun Liu\thanks{Shanghai Jiao Tong University. Email: liuliqunallen@gmail.com.}}
	\maketitle
	\begin{abstract}

Motivating careerists is challenging for political organizations. Without explicit contracts, careerists often pander to public opinions or their superiors' preferences. Worse, when tasked with implementing these distorted decisions, they tend to underinvest in the necessary efforts.  We analyze the motivation problem by  examining how a careerist agent  fulfills these roles on behalf of a principal across various information structures. Importantly, the principal can credibly commit to  performance-based reward schemes to incentivize correct decisions and diligent implementation. However, such schemes are feasible only if the principal observes policy consequences while backing away from implementation details. Along the way, we characterize the principal-optimal information structure. Putting theoretical findings into practice, we explore the underlying incentive structures and their policy implications.

   	\vspace*{0.5cm}

  \noindent \textbf{Word count:} 7483
\end{abstract}

\clearpage
\pagenumbering{arabic}

\section{Introduction}
Motivation is crucial for the effective operation of organizations. 
In economics, motivation is typically achieved through explicit contracts. In politics, such contracts are rare. Instead, careerists often align their decisions  with public opinion or the preferences of their superiors (e.g., \cite{maskin2004politician}, \cite{fox2007government}). This can not only distort decision-making, but also demotivate those responsible for carrying out the decisions they have made.

Two examples help illustrate this point. Within parliamentary systems, department aides can become frustrated when pressured to both agree to and effectively carry out tasks they do not support. The appointment of Dominic Cummings as a senior advisor to Boris Johnson, a key Brexit advocate, notably demotivated many British civil servants. Cummings equated support for a no-deal Brexit with loyalty. He promised extra Treasury funds to department aides, intending to compel them to actively support and implement the Brexit agenda. Essentially, this created a ``do or die'' scenario for department aides.\footnote{``How Dominic Cummings took control in Boris Johnson’s first days as Prime Minister''. {\it BuzzfeedNews}. 27 July 2019.}   But not all civil servants shared his enthusiasm for Brexit; in fact, many felt confused and demoralized. For example, after grappling with the complexities of Brexit for an extended period, one civil servant lamented to {\it The Guardian}, ``Heaven help us if no deal (Brexit) actually happens.''\footnote{``Many civil servants are depressed – including me. Brexit will do that to you''. {\it The Guardian}, 26 November 2019.} Former treasury official Jill Rutter described Brexit as ``an article of faith.''\footnote{ ``The civil service must speak truth to Boris (and his Cabinet).'' The Institute for Government, 25 July 2019.}  Dave Penman, head of the FDA union, criticized the enforcement of unproven ideologies that could negatively impact public service delivery.\footnote{``Dominic Cummings's role provokes alarm inside civil service.'' The Guardian, 25 July 2019.}

Within hierarchical government structures, demotivation often arises when lower-tier officials pander to their superiors' policy preferences in pursuit of political rewards. This leads to a phenomenon known as  ``face innovation,'' where Chinese local officials launch reforms primarily to appear progressive \citep{teets2015politics}. However, these reforms often lack effective implementation, leading to unsustainable innovations and resource misallocation. One newspaper sharply criticized such practices, stating: ``these projects serve not the public good but personal fame and recognition. They often result in flashy but superficial undertakings that waste resources and harm rather than benefit the people.''\footnote{``Targeting Formalism `Face Projects.' '' Dongfang Net, 10 December 2019. For related articles, see also ``Innovating for the Sake of 'Innovation' is a Form of Formalism.'' Zhongguo Jiwei Jiancha, 3 November 2020.}

In summary,  political careerists often feign compliance to secure their positions when given the option to change the status quo.  The absence of credible incentive structures—for example, neither Cummings nor the Chinese central government can write complete contracts tying rewards to the policymaking process and outcomes for these careerists—leads to distorted decisions and implementation. Addressing such motivation problems require political rather than merely economic solutions.

We ask: in situations where careerist agents are responsible for both choosing and implementing  policies, under what conditions are they fully dedicated to pursuing the best policy outcomes? How can their performance be credibly rewarded or penalized?  Furthermore, when is it preferable to delegate policymaking to a careerist agent over a non-careerist? Lastly, under what conditions should the principal  reclaim policymaking authority rather than delegate?

To address these questions, we pursue a novel extension of the classical pandering models (e.g., \cite{maskin2004politician, fox2007government}). As is standard, our model involves an informed agent delegated the task of policymaking on behalf of an uninformed principal; the principal is uncertain about whether the agent has aligned  policy preferences; at the end of the policymaking process, the principal evaluates the available information and decides whether to retain or replace the agent.

Our model features two novelties. Firstly, unlike most existing literature  where policymaking is  one-dimensional, ours introduces a multidimensional policymaking space. The agent first chooses between maintaining the status quo or pursuing reform, and then decides the costly effort needed for implementation. This scenario is particularly relevant when policymaking power is concentrated in a single entity accountable to constituents or superiors, such as local officials in China who adapt policy innovations from other regions based on local conditions (e.g., \citet{heilmann2008local}). Reform is riskier than the ``safe'' status quo, as it can fail due to bad decisions or poor implementation. Failed reforms, as seen in many policy innovations, can have severe consequences. In contrast,  Lee's model of legislative process assumes that a failed reform simply reverts to the status quo \citep{lee2024}.

Secondly, although the principal in our model is not directly engaged in the policymaking process, she can select varying levels of ``transparency'' to oversee the agent's actions. Given the expanded dimension of policymaking, we accordingly modify the notions of transparency from existing literature (e.g., \cite{prat2005wrong}, \cite{fox2012costly}).  We assume throughout that the policy choice is observable to the principal, reflecting the realistic scenario where any significant reform decision or policy deviation from the status quo cannot easily be hidden from the public. This leads to three relevant transparency levels in reform policymaking: \textit{Non-Transparency}, where the policy choice is observable to the principal; \textit{Semi-Transparency}, where the policy choice and its outcomes are  observable to the principal; \textit{Full-Transparency}, where the policy choice, its detailed implementation, and outcomes are all  observable to the principal.\footnote{In the Appendix, we also explore the implication of other transparency levels. }

We examine the decision-making and implementation strategies of a careerist agent under varying levels of transparency.  Our key finding suggests that an intermediate level of transparency effectively resolves the commitment problem in the principal-agent relationship. At this level, the principal can credibly commit to electorally rewarding or punishing the agent contingent on policymaking outcomes (though not on implementation effort, as it remains unobserved).  Under such an incentive scheme,   successful policymaking leads to both political and policy rewards, while failure leads to punishments. This dual benefit provides  the agent with high-powered incentives to exert effort in implementation. Conversely, when transparency is either too high or too low, the agent focuses on merely acting congruently  rather than genuinely pursuing successful policymaking.

To understand why an intermediate level of transparency enables the principal to commit to a performance-based electoral scheme, we consider how the principal updates her beliefs about the agent's alignment. For such a scheme to work, the principal must positively update her beliefs when outcomes are good, and negatively when they are bad. This scheme is not feasible under minimal transparency where outcomes are unobservable (\textit{Non-Transparency}). With excessive transparency (\textit{Full-Transparency}), chances are that the agent implements so diligently in the policymaking process that the principal is almost convinced of his alignment, yet the project fails due to idiosyncratic reasons. In such cases, a forward-looking principal should reelect the agent despite bad outcomes, which contradicts the notion of a ``performance-based'' reward scheme. Between the extremes where implementation details are concealed but outcomes are visible, the principal infers alignment solely from these outcomes. Since aligned agents typically have a stronger motivation to deliver successful policies and often do so,  success signals alignment. Therefore, a performance-based electoral scheme is most likely under this intermediate level of transparency.

We also explore the welfare implications of different transparency levels. \textit{Non-Transparency} always results in the lowest welfare for the principal. This finding resonates with previous literature, where ``transparency on action'' leads careerist agents to feign rather than choosing the optimal policy \citep{prat2005wrong,fox2007government}. More importantly, we identify scenarios where \textit{Semi-Transparency} might outperform \textit{Full-Transparency}. While \textit{Semi-Transparency} guarantees a performance-based reward scheme that links political and policy rewards to motivate implementation efforts, \textit{Full-Transparency} may motivate careerists to work diligently for reelection by meeting the  threshold endogenously set for this purpose. Ex ante, the comparison between these levels of transparency remains uncertain.

The welfare implications are particularly relevant in real-world scenarios where a principal, unable to write explicit contracts, opts to enhance oversight to better discipline an agent. For example, a central government might establish monitoring bodies (such as a supervisory department) to oversee local officials' policy implementation beyond merely offering promotion incentives. For example, in a university where ``satisfying academic performance'' ambiguously defines tenure criteria, a dean might add annual reviews to the existing 3-year reviews to better monitor assistant professors' progress. While these practices allow principals to gather more detailed data on processes rather than outcomes, our theoretical analysis suggests that they could actually be counterproductive.

Our study contributes to the debates about the right kind of transparency. \cite{prat2005wrong} shows that knowing  the agent's actions in addition to policy outcomes can sometimes hurt the principal. \cite{fox2012costly} demonstrate that with highly imbalanced priors, the principal benefits from backing away from full transparency. Both results hinge on the ``invertibility'' from  outcomes to actions.
That is, conditional on observing policy outcomes, the principal always has more to learn about the agent's types by observing the actions. This assumption may not hold in the context of reform policymaking, where a principal can infer whether a reform was implemented based on the observable outcomes—success, failure, or the status quo. Given that observing outcomes generally matters for the principal, the key question is whether also observing the implementation process helps incentivize better policy choices and implementation. Our results revisit the role of transparency along this line, and identify situations where knowing more may hurt.

In a related paper, \cite{ashworth2014voter} demonstrate that  voters do not necessarily benefit from observing the policy choice of incumbents if they have observed the policy outcomes. While our findings similarly suggest that more information is not always a blessing, our approaches and model scopes differ. Firstly, our model operates within a multidimensional space, unlike their one-dimensional space. This allows our analysis to more naturally apply to the context of reform policymaking. Secondly, although our results may appear superficially similar, they fundamentally diverge. Our study focuses on varying levels of transparency to identify the optimal types of information to gather. We find that when a principal can observe both the policy choice and its implementation, outcome data may become irrelevant for electoral decisions. Thus, optimal transparency is not merely about gathering more information but about identifying the most relevant information.

Our study further contributes to the theory of delegation. The ``ally principle'' of classical delegation theory  argues that a less informed principal should delegate decision-making rights to a more informed  yet potentially biased agent only if the benefits derived from that agent's information exceed the loss of control (e.g., \cite{holmstrom1980theory}, \cite{bendor2001theories}). In an extension of our model, our results lend further support to this theory within a more complex policymaking environment. By selecting an intermediate level of transparency during delegation, the principal can motivate the agent to make informed decisions and exert considerable implementation effort, thereby minimizing the adverse effects of losing control over a misaligned agent.

\cite{maskin2004politician} suggest that it could be beneficial to delegate policymaking to nonaccountable officials (those exempt from reelection concerns) as opposed to their accountable counterparts. Central to their argument is the idea that career concerns, such as reelection incentives, could demotivate accountable officials  from gathering information essential for making informed decisions. 
Contrary to this, our extension of the baseline model suggests that accountable officials can actually be more motivated than nonaccountable ones, given the appropriate level of transparency. When these accountable officials implement policies, they are guided not only by the pursuit of leaving a policy legacy, but also by the allure of a promising career.

Our theoretical model builds on the pandering literature. A key theme along this line is that  career concerns can lead agents to ignore valuable signals  to  appear smart \citep{canes2001leadership,levy2005careerist,prat2005wrong,fox2012costly} or loyal \citep{maskin2004politician, fox2007government,fox2009delegates,fox2011delegation}. Such ``pandering'' behaviors can distort policy choices, ultimately reducing the principal's welfare. Our study's technical contribution involves delving deeper into the agent's behavior within an expanded dimensionality of policymaking. In this setting, determining the optimal level of transparency that aligns decision-makers' political incentives with effective policymaking becomes increasingly challenging.

In many agency problems,  the principal benefits from gaining more information about the agent's actions (e.g., \cite{holmstrom1979moral}). However, several studies have predicted the opposite. For instance, \cite{holmstrom1999managerial} illustrates that acquiring more accurate information about the agent could reduce the agent's motivation to exert effort. Similarly, \cite{cremer1995arm} argues that more information might adversely affect policymaking when considering renegotiation possibilities. \cite{dewatripont1999economics} argue that agents might work harder when the principal receives a noisier signal about performance. Departing from the commitment assumption common in most contractual situations, our study adds to this body of work by suggesting that more information can be detrimental, particularly because the agent may lack the proper motivation to pursue the right objectives. In a related vein, \cite{bdm2015political} show that no transparency combined with a flexible budget can mitigate inefficiencies in a dynamic moral hazard setting.

\section{Model}
\subsection{Policy setting} A privately informed careerist agent (referred to as ``he'')  chooses and implements a policy on behalf of a principal (referred to as ``she''). The policy $x$ can either be the status quo policy ($x=q$) or a reform ($x=r$). The status quo leads to  a sure policy outcome. The reform outcome can be ``successful'' or ``unsuccessful,'' which depends on its state of nature $\omega\in \{G,B\}$ and the implementation effort $e\in [0,1]$. Here $\omega=G$ means that a reform is good by nature and $\omega=B$ means that a reform is bad by nature. The prior distribution of $\omega$ is $P(\omega=G)=\phi\in (0,1)$. We maintain the standard assumption that the agent observes $\omega$, while the principal does not.  

For a successful reform, ``choosing well'' and ``implementing well'' are complementary: a bad reform always fails, while a good reform succeeds with a probability exactly equal to the implementation effort $e$ chosen by the agent.\footnote{Alternatively, we can think of the agent directly choosing the probability that a good reform shall succeed,  with higher costs as the probability increases.} The agent incurs a private implementation cost $c(e)=\frac{e^2}{\lambda}$. Depending on the information structure (to be discussed later), the effort $e$ can be either made public or concealed. For example, in a ``transparent'' environment, politicians reveal how actively they have promoted an innovative economic policy through media coverage; in an ``opaque'' one, they conceal this information.  $\lambda$ parameterizes the agent's cost sensitivity to the effort, with a larger value of $\lambda$ indicating a lower implementation cost.

The setting applies broadly to situations where the agent's role is more than choosing a policy:
\begin{enumerate}
\item \textit{Experimentation within federal systems}. A reform opportunity arrives. Based on prior trials, the central government (principal) believes each state  has a success probability of $\phi$ conditional on perfect implementation. A career-concerned local governor (agent) privately learns whether the reform adapts to local conditions ($\omega$). He chooses both whether to reform and the extent to which it should be implemented.
\item \textit{Integrated decision-making and monitoring, divided implementation}. Reelection-seeking politicians create policies while bureaucrats carry them out (e.g., \cite{fox2011delegation}). In addition, politicians invest effort $e$ in overseeing bureaucratic implementation. More oversight results in better implementation, but it also imposes a higher private cost on politicians.

\item \textit{Partially publicizable implementation}.  Instead of the entire implementation process, only specific milestones (e.g., midterm reviews) of the implementation can be made public. A publicizable milestone $e$ acts as an ``intermediate product'' governing the success probability through a known increasing function $g(e)$. We relabel the success probability $\Tilde{e} = g(e)$ as the ``final product''  and treat it as the agent's actual choice variable with an associated cost $c(\Tilde{e})=c(g(e))$. The analysis proceeds analogously by treating $\Tilde{e}$ as $e$ in the baseline.
\end{enumerate}

\subsection{Strategies}
The agent has a private type $t\in \{C, N\}$ governing his policy preference.  With probability $\pi\in (0,1)$ the agent is a congruent type ($t=C$) whose policy preference is aligned with the principal: they both prefer a successful reform to the status quo to a failed reform. With probability $1-\pi$ the agent is a noncongruent type ($t=N$) who prefers maintaining the status quo to any reform outcome. Consistent with \cite{fox2007government}, we define noncongruence as an agent who resists policy changes due to intrinsic biases or external influences from interest groups.\footnote{In Section \ref{R5:congruence} of the Appendix, we further elaborate on this notion of noncongruence.} 

Both types of the agent aim to appear congruent by the end of the policymaking process. This  reputation for congruence is valued because it can lead to political rewards from the principal. The nature of these rewards can vary depending on the principal's objectives, such as identifying a loyal ally or partnering with an ``open-minded'' implementer for future reforms. We model this reputation payoff as influencing the principal's retention decision. If the agent is not retained, the principal will select a replacement randomly from the same pool of candidates, where each candidate has a probability $\pi$  of being congruent.

There are four possible ``payoff types'' of the agent, denoted by the pair $(t,\omega)\in \{C, N\}\times \{G,B\}$. Hereafter, we use ``type $(t,\omega)$'' to denote a type $t$ agent who has observed state $\omega$. Formally, the agent's strategy maps the private type into the distribution of actions. Thus, his strategy can be written as $(\kappa (t,\omega),e(t,\omega))\in [0,1]\times [0,1]$, the probability  with which an agent initiates a reform and the associated implementation effort for each type $t\in \{C,N\}$ and the state $\omega\in \{G,B\}$. We denote $e(t,\omega)=0 $ if $\kappa(t,\omega)=0$, implying that a type $(t,\omega)$ agent who does not reform exerts no implementation effort. The principal forms beliefs about the agent's private type $t$ using information available to her in an environment $J$ (which shall be specified later). Her (mixed) strategy maps from the posterior belief $\mu^J$ that the current agent is congruent to the probability of retention. 
\subsection{Payoffs}
The agent's utility function depends on the outcome of policymaking and retention.  Let $u_t(x,e,\omega)$ be a type $t$ agent's policy payoff, and $R$ be his value of retention. Then a type $t$ agent's utility function can be written as $ U_t(x,e,\omega)=u_t(x,e,\omega)-c(e)+R\cdot \mathbbm{1}\{\text{retention}\}$,
where 
\begin{align*}
   u_C(x,e,\omega)=\begin{cases}
       0 \quad &\text{if} \quad x=q\\
       -v \quad &\text{if} \quad \text{reform fails}\\
              v \quad &\text{if} \quad \text{reform succeeds}
   \end{cases},
  u_N(x,e,\omega)=\begin{cases}
       0 \quad &\text{if} \quad x=q\\
       -v \quad &\text{if} \quad \text{reform occurs}
        \end{cases}      .
\end{align*}
The policy payoff structure ensures that the noncongruent agent has a clear preference for the status quo over the reform decision. Conversely, the congruent agent is penalized for either selecting a bad reform or inadequately implementing a good one; he may be rewarded only if a good reform is correctly chosen and effectively executed.

Notably, the congruent agent's payoffs from reform outcomes are symmetric around zero, the payoff for maintaining the status quo. We do so to simplify computations and emphasize the motivation issue in the agency relationship. In the Appendix (Lemma \ref{lemma:bad incentive}), we show that without career incentives, an agent's optimal choice is to maintain the status quo. Reform may be chosen only if the agent has been sufficiently motivated to pursue successful reform policymaking. Thus, our setup illustrates the conditions under which strong career incentives are provided. In Section \ref{R2: asymmetric} of the Appendix, we relax the symmetric assumption and find our core insights intact.

The principal's utility function depends on policymaking outcomes and political selection. Let $u(x,e,\omega)$ be the principal's policy payoff, and $Q$ be the probability that the selected agent is congruent after the policymaking process.\footnote{If the current agent is retained after the policymaking, $Q$ equals the principal's posterior belief that the current agent is congruent.  If the current agent is replaced,  $Q$ equals $\pi$.} Then the principal's utility can be written as $U(x,e,\omega)=u(x,e,\omega)+\alpha \cdot Q$, where the policy payoff takes the form 
\begin{align*}
   u(x,e,\omega)=\begin{cases}
       0 \quad &\text{if} \quad x=q\\
       -v \quad &\text{if} \quad \text{reform fails}\\
              v \quad &\text{if} \quad \text{reform succeeds}
   \end{cases}.
\end{align*}
Here, the parameter $\alpha>0$ represents the principal's net benefit from choosing a congruent agent as opposed to a noncongruent type. The interpretation of $\alpha$ can vary based on the context. For example, a reformist principal may incur a partisan loss of 
$\alpha$ when working with a conservative agent as opposed to a reformist one. Alternatively, an open-minded agent might support a reform project valued at  $\alpha$ by the principal, while a conservative one would not.

\subsection{Information structures}
Our analysis focuses on comparing the agent's behavior and the quality of policymaking across various environments, with a particular emphasis on three distinct information structures. These structures are organized by their levels of transparency.

We begin by describing the setting in which the principal observes only the reform decision before evaluating the agent. We refer to this environment as \textit{Non-Transparency} (\textit{NT}). The agent's congruence is evaluated by
\begin{align*}
    \mu^{NT}(x):=P(t=C|x), \qquad x\in \{r,q\}.
\end{align*}

With \textit{Non-Transparency},  the principal's retention decision is made before the implementation details and consequences of policymaking become available. This is a familiar scenario in literature\footnote{See \citet{canes2001leadership,maskin2004politician,fox2012costly}.} and real-world situations, such as a politician facing reelection before policy outcomes are realized.

Next, we describe scenarios in which the principal observes policy consequences in addition to the reform decision. Depending on whether the agent's effort is observable, these environments can be further divided into \textit{Semi-Transparency} (\textit{ST}) and \textit{Full-Transparency} (\textit{FT}). This classification extends \cite{prat2005wrong}'s notion of ``transparency on consequence'' to more complex policymaking that includes both policy choice and implementation.

With  \textit{Semi-Transparency}, the principal observes both the reform decision and the outcomes before deciding on retention. The agent's congruence is evaluated by
\begin{align*}
    \mu^{ST}(x,y):=P(t=C|x,y), \qquad x\in \{r,q\},y\in \{q,S,F\}.
\end{align*}
where $S$ stands for ``successful reform'' and $F$ stands for ``unsuccessful reform.'' 

Compared to \textit{Non-Transparency}, \textit{Semi-Transparency} enables the principal to condition the retention decision on observed policy outcomes. A real-world interpretation of \textit{Semi-Transparency} could involve a situation where voters can see the policy outcome, but secrecy rules or limited oversight prevent them from observing the implementation process. For instance, in diplomatic efforts, voters may know the final terms of a treaty or agreement, but the details of the negotiation process and specific actions taken during diplomatic talks remain hidden from public view.

With \textit{Full-Transparency}, the principal observes every aspect of policymaking, including the reform decision, implementation effort, and the consequences. The agent's congruence is evaluated by
\begin{align*}
    \mu^{FT}(x,y,e):=&\begin{cases} P(t=C|r,y, e), x=r,y\in \{S,F\}, e\in [0,1]\\
    P(t=C|q), x=q
    \end{cases} ,
\end{align*}

\textit{Full-Transparency} represents an environment where the principal has the capacity for strict monitoring over the agent. For example, voters may witness the execution of a policy through social media, observing not just the final outcome but the entire process. Similarly, higher-ranking officials responsible for promotions may monitor all aspects of the actions taken by subordinates via monitoring bodies.

\subsection{Timing of interactions}
The game proceeds as follows:
\begin{enumerate}
    \item Nature picks the random variables $(\omega,t)$ according to the prior distributions. 
    \item The agent observes $\omega,t$, and chooses $x\in\{r,q\}$. Conditional on $x=r$, he chooses implementation effort $e\in[0,1]$.
    \item  The policy consequence $y$ is realized. 
    \item The principal decides whether to retain the agent conditional on her available information. 
    \item All players' payoffs are realized.
\end{enumerate}

\subsection{Interpretation}\label{IA}
One interpretation of the model is that reelection-seeking politicians signal their desirable attributes  with policymaking (e.g., \citet{canes2001leadership,maskin2004politician,fox2007government}). Our model expands on this foundational setup by considering scenarios where policy outcomes are influenced not only by the quality of decisions but also by the efficacy of implementation, both of which are the responsibility of the same politician. The key focus is on how to strategically use varying levels of transparency to incentivize better policymaking by these politicians.

More broadly, the model applies to settings beyond electoral politics. The key characteristic is that within political agency, the agent's future career hinges on the reputation built from  multidimensional decision-making. In our motivating examples, department aides must agree to and implement Brexit diligently to secure their positions, with Dominic Cummings as the principal and the aides as agents. Similarly, in hierarchical government structures, the central government—acting as a unitary actor responsible for promoting local officials—serves as the principal, while local politicians are the agents. These local officials are given the option to pursue reforms and are responsible for implementing them, with promotion as the political reward.

Before proceeding to the analysis, we impose two assumptions. Firstly,  $\lambda(v+\frac{R}{2})\leq 1$. This technical restriction ensures that $e$ falls within $[0,1]$, allowing $e$ to represent the probability of a successful reform. Secondly, to focus on the interesting cases, we assume that the value of retention ($R$) must not be too small. If it is, the agent may simply choose their preferred policies without concern for their future career. Formally, we require that $R \geq \max\{2\sqrt{\frac{v}{\lambda}},v\}$, and discuss its implication in Section \ref{R1: office} of the Appendix. The parameters $(R,\lambda,v)$ that satisfy these assumptions are non-degenerate in the parameter space.

\section{Analysis}
\subsection{Definitions}

The solution concept of the game is the Perfect Bayesian Equilibrium (PBE). This requires the principal to form beliefs about the agent's type $t$ using the Bayes rule wherever possible, and to act in a manner that respects sequential rationality. Since PBE does not restrict players' off-path beliefs, we exclude ``unreasonable'' equilibria that do not meet the universal divinity criterion from \cite{banks1987equilibrium}. Simply put, this criterion requires that any unexpected action should come from the agent type most likely to benefit from such an action. However, the divinity criterion cannot rule out an uninteresting ``inaction equilibrium'' across all information structures,  where all types of the agent maintain the status quo and are retained along the path.\footnote{The formal statement and proof is in the Appendix (Lemma \ref{Eq: inaction}). This inaction equilibrium is uninteresting, as it masks the influence of information in reform policymaking. }  Thus, we look for equilibria that do not feature inaction. We also restrict attention to equilibria where the principal employs a pure retention strategy, which is standard in the literature (e.g., \cite{maskin2004politician,armstrong2010model}).

Three potential equilibria exist that are exhaustive and mutually exclusive. An equilibrium is \textit{unresponsive} if all types of the agent choose the same policy. An equilibrium is \textit{responsive} if the policy choices of both congruent and noncongruent agents vary with the state $\omega$. For example,  an equilibrium where a type $(t,G)$ agent reforms while a type $(t,B)$ agent maintains the status quo is responsive. Finally,  an equilibrium is \textit{semi-responsive} if the policy choice of 
exactly one of the congruent and the noncongruent agent remains constant across the state $\omega$. An example of this would be a scenario where a congruent agent chooses to reform whenever $\omega=G$ but a noncongruent agent always maintains the status quo.

\subsection{Equilibria}
\begin{Proposition}[Non-Transparency]\label{Prop:NT} Under Non-Transparency,
 an unresponsive equilibrium exists and is unique:  all types of the agent choose to reform and are retained along the path. Further, a type $(C, G)$ agent implements with effort $\lambda v$; other agents implement with effort $0$.
\end{Proposition}

Under  \textit{Non-Transparency}, the strong motive to hold office drives all types of the agent to pander to the ex ante popular ``reform'' decision without utilizing their information expertise. In equilibrium, the principal's decision to retain the agent hinges solely on whether a reform decision is observed. As such,  the principal cannot learn about the alignment of the agent through their policy choices. This equilibrium  resembles the ``full pandering'' scenario described in \cite{maskin2004politician}, where strong office-holding motives distort policymaking. Since the principal cannot condition her retention strategies on unobserved implementation details or policy outcomes, agents are incentivized to underinvest in policy implementation to reduce costs.

\begin{Proposition}[Semi-Transparency]\label{Prop:ST} Under Semi-Transparency,
a responsive equilibrium exists and is unique: a type $(t,G)$ agent reforms, while a type $(t,B)$ agent maintains the status quo. 
Further, whenever a reform occurs, 
   a congruent agent implements with effort $\lambda (v+R/2)$ while a noncongruent agent implements with effort  $\lambda R/2$.  Retention occurs if only if a reform succeeds.
\end{Proposition}

Under \textit{Semi-Transparency}, the observability of policy outcomes  allows for the possibility that the principal may leverage them to provide incentives. For such incentive schemes to be credible, it must be in the principal's interest to retain the agent following successful reforms and replace them following unsuccessful ones. This requires the principal to positively update about the agent's alignment following successful reforms, and negatively following unsuccessful ones. A congruent agent, motivated not only by retention but also by the intrinsic reward of successful reforms, invests greater implementation effort and succeeds more often than a noncongruent agent. Thus, a rational principal will update her beliefs accordingly, thereby justifying this retention strategy.

With the current incentive scheme in place, a successful reform offers the agent both policy and political rewards. Given high stakes, both congruent and noncongruent agents are motivated to reform when the conditions  are favorable. When the conditions are unfavorable, the agent will  maintain the status quo anticipating potential replacement. He will not switch to a reform decision likely to fail, as it would also result in his replacement. Therefore, under \textit{Semi-Transparency}, the agent's policy choice is entirely responsive to the conditions of the reform, which ensures the quality of the reform.

While  the incentive scheme under \textit{Semi-Transparency} can effectively  motivate implementation, they are inherently imperfect. To encourage reform decisions, the principal must sometimes electorally punish those who correctly maintain the status quo. This is the political cost the principal must bear—potentially punishing someone who made the right decision. If she fails to do so, the misaligned type may exploit the situation by choosing the status quo and reaping political benefits even when conditions would favor reform.

\begin{Proposition}[Full-Transparency]\label{Prop:FT} 

Under Full-Transparency, 
\begin{enumerate}
    \item  if $R<\lambda v^2+v$, a semi-responsive equilibrium exists and is unique. Specifically, a type $(C,G)$ reforms with implementation efforts $\lambda v$ and is retained; other agents  choose the status quo and are replaced.
    \item  if $R\geq \lambda v^2+v$, a semi-responsive equilibrium exists: a type $(C,G)$ reforms with implementation efforts $\sqrt{\lambda(R-v)}$ and is retained; other agents  choose the status quo and are replaced. Moreover, a continuum of unresponsive equilibrium exists: all types of the agent choose to reform with implementation effort $e^*\in [\lambda v,\sqrt{(R-v)}]$ and are retained along the path.
\end{enumerate}
\end{Proposition}
Under \textit{Full-Transparency}, two kinds of equilibrium may arise. With weak office-holding motive ($R<\lambda v^2+v$), a type $(C,G)$ agent stands out from other types by launching a reform. With strong office-holding motive ($R\geq \lambda v^2+v$), every agent type may reform. While these equilibria echo those seen under \textit{Non-Transparency}, a difference emerges: as policymaking becomes more transparent to the principal, it gets harder for different types of the agent to pool by pandering to the congruent action that ensures retention. This is highlighted by the equilibrium threshold: under \textit{Full-Transparency}, we need $R\geq v^2+v$ in addition to the assumptions  in Section \ref{IA} to attain a  pooling (unresponsive) equilibrium, whereas such is not the case with \textit{Non-Transparency}.

Notably, a continuum of unresponsive equilibria survives the divinity criterion when office-holding motives are strong. The intuition behind this equilibrium mutiplicity is that if all types of the agent  choose to reform with the same equilibrium implementation effort and are guaranteed retention, increasing efforts is unprofitable. However, reducing efforts would imply misalignment and cause job loss.

Despite the issue of equilibrium multiplicity, across all types of equilibria an agent's future career hinges on the actions chosen rather than their induced outcomes. Ideally, the principal would write a contract specifying retention strategies that potentially vary with both implementation efforts and policy outcomes. However, as policy outcomes become decoupled from retention decisions under \textit{Full-Transparency}, the agent's motivation may weaken.

Before proceeding, it is important to compare the agent's strategic behaviors as transparency levels vary. When  career concerns are strong, policymakers tend to opt for ex ante popular actions rather than selecting policies that are fundamentally appropriate (e.g., \cite{canes2001leadership,maskin2004politician,fox2007government}). Such ``pandering behaviors'' prevent policymakers from adapting policies to the 
actual state of the world, thus reducing the quality of policymaking. A commonly proposed solution to this pandering phenomenon is to back away from full transparency \citep{ashworth2012electoral}. Yet, the relationship between pandering and transparency seems nuanced, especially in the context where pivotal decisions like reform policymaking cannot be conveniently concealed from the public's eye. We demonstrate with Propositions \ref{Prop:NT}-\ref{Prop:FT} that policymakers' pandering incentives do not straightforwardly rise or fall with increasing transparency levels.  When career concerns are particularly strong ($R\geq \lambda v^2+v$),  all agent types may resort to pandering towards congruent actions under both \textit{Non-Transparency} and \textit{Full-Transparency}, but not in the case of \textit{Semi-Transparency}.

Comparing Propositions \ref{Prop:NT}-\ref{Prop:FT}, we see that more transparency generally improves sorting. This is because agents intrinsically averse to reform decisions find it more costly to mimic others when more actions are observable (see also \citet{cremer1995arm, dewatripont1999economics, holmstrom1999managerial}). However, this pattern is not absolute. When career incentives are particularly strong ($R\geq \lambda v^2+v$), all types may pool on the same behaviors under both \textit{Non-Transparency} and \textit{Full-Transparency}, thereby preventing sorting. In contrast, \textit{Semi-Transparency} allows the principal to achieve some degree of sorting.

\subsection{Welfare Comparison}
\textit{Non-Transparency (NT)}. -- Under \textit{Non-Transparency}, all types of the agent pander to the ex ante congruent policy ``reform.'' Only the type $(C,G)$ agent may succeed, as he implements with effort $\lambda v$ when the state calls for a reform. In such an equilibrium, the principal learns nothing about the agent. Thus, the principal's expected welfare is 
\begin{align*}
    W^{NT}=2 \pi \phi \lambda v^2-v+\alpha \pi .
\end{align*}

\textit{Semi-Transparency (ST)}. -- Under \textit{Semi-Transparency}, all types of the agent pursue successful policymaking. A reform decision happens when the agent observes a good reform opportunity, which is implemented with effort $\lambda (v+\frac{R}{2})$ by the congruent agent  and   $\lambda \frac{R}{2}$ by the noncongruent agent. Otherwise, the status quo is maintained. In this equilibrium, the principal learns nothing about the agent if the state is $\omega=B$. When the state is $\omega=G$, she retains a congruent agent with probability $\lambda (v+\frac{R}{2})$ and retains a noncongruent agent with probability $\lambda \frac{R}{2}$. 
Thus, the principal's expected policy payoff is 
\begin{align*}
    W^{ST}= 2 \pi \phi \lambda v^2+\phi \lambda Rv-\phi v+\alpha \pi[1+(1-\pi)\phi \lambda v ].
\end{align*}

\textit{Full-Transparency (FT)}. -- Under \textit{Full Transparency},  depending on parameters (whether $\lambda v^2\leq \lambda (R-v)$ holds), two kinds of equilibrium may exist. 
It is straightforward to verify that the principal's payoff takes the following forms: 
\begin{align*}
    W^{FT}=\begin{cases}
        2\pi \phi \max \{\sqrt{\lambda (R-v)}, \lambda v\}v -\pi \phi v+\alpha \pi [1+(1-\pi)\phi]. (\textbf{SS}) \\
         2 \phi e^* v- v+\alpha \pi, e^*\in [\lambda v, \sqrt{R-v}]. (\textbf{US})
    \end{cases}
\end{align*}
where \textbf{SS} refers to the semi-responsive scenario, while \textbf{US} refers to the unresponsive scenario.  

Now, we are in a good position to compare the principal's welfare under different information structures. For expository convenience, parameters $(R,\lambda,v)$ are said to satisfy ``\textit{Condition M}'' if $\lambda v^2\leq R-v,\lambda v<1/3$ and $R\in (\underline{R}, \Bar{R})$, where $\underline{R}=\frac{1-\lambda v-\sqrt{1-3\lambda v}}{\lambda /2}$ and $\overline{R}=\frac{1-\lambda v+\sqrt{1-3\lambda v}}{\lambda /2}$. 

\begin{Proposition}[Comparison of information structures]\label{Prop:ID} For the principal, Non-Transparency always induces the lowest welfare. Full Transparency is preferred to Semi-Transparency if one of the following holds:
    \begin{enumerate}
        \item the semi-responsive equilibrium is selected, and either (1) $\alpha$ is large, or (2) the condition M holds and $\pi$ is sufficiently close to $1$. 
        \item the unresponsive equilibrium is selected, the condition M holds, and the equilibrium $e^*$ is sufficiently close to $\sqrt{\lambda (R-v)}$. Furthermore, $\phi$ is sufficiently close to $1$ and $\alpha$ is small. 
    \end{enumerate}
    In all other situations, Semi-Transparency is preferred. 
\end{Proposition}
Proposition \ref{Prop:ID} says \textit{Non-Transparency} is strictly dominated, as it leads the agent to choose reforms without serious implementation efforts. The principal should decide between \textit{Semi-Transparency} and \textit{Full-Transparency}.  Part I suggests if $\alpha$ is large, the principal should increase transparency to more effectively screen the agent. This result replicates the well-known results that more transparency improves sorting (e.g., \cite{cremer1995arm, dewatripont1999economics,holmstrom1999managerial}). 

To gain more insights into the comparison, we suppose that  the principal's top priority is policymaking ($\alpha$ is small). A critical question then arises: Does a type $(C,G)$ agent, the type most dedicated to a reform, exert more implementation efforts under \textit{Semi-Transparency} than \textit{Full-Transparency}? The answer is indeterminate. Under \textit{Full-Transparency}, such types focus on diligent action; under \textit{Semi-Transparency}, they aim for successful reform outcomes. We provide conditions under which each information regime might be more effective. If type $(C,G)$ agents are sufficiently common, it is unclear whether the principal necessarily benefits more from \textit{Semi-Transparency} or \textit{Full-Transparency}. In essence, more information does not lead to better outcomes. 

We connect our findings to the debate on whether ``transparency on action'' or ``transparency on consequence'' improves policymaking quality. \cite{prat2005wrong} shows that when only policy consequences are observed, careerist agents are incentivized to choose intrinsically good policies. However, when actions are also observed, these agents may herd on priors rather than utilizing their own information. In such cases, more transparency can actually harm the policymaking process. \cite{fox2012costly} further highlight that this result may depend on the cost structures of policy mistakes. In both works, the agent's action space is one-dimensional, and the principal’s policy payoff depends solely on the action taken.

Matters are different when the dimensionality of policy expands. 
First, ``transparency on action'' should also include implementation in addition to policy choices. The principal can  infer whether a reform occurred based on three possible outcomes: ``successful reform,'' ``unsuccessful reform,'' or ``status quo.'' Thus, transparency on consequences implies transparency on reform decisions. Second, since actions (policy choice and implementation) govern the distribution of policy consequences, knowing the outcome becomes redundant for inferring the agent's type if the principal already knows both the policy choice and the implementation details.

Our equilibrium analysis, in light of \cite{prat2005wrong} and \cite{fox2012costly}, can be decomposed into solving several questions: (1) Is the agent more likely to herd on popular decisions when the principal observes the implementation, assuming the outcomes are already observable? (2) Is the agent less likely to herd when the principal observes either the implementation or the outcomes, assuming the policy choice is observable? (3) Most importantly, once a reform is initiated, does the agent put in more implementation effort when more information is observable?

The answers to questions (1) and (2) echo prior research: observing outcomes can discipline the agent to make the right decisions, while observing more than that can incentivize the agent to make conformist decisions of reform.  
However, this alone is insufficient to address more complex policymaking where the quality of implementation also matters. \textit{Full-Transparency} might be worse than \textit{Semi-Transparency} in incentivizing the correct reform decisions. Yet, once a reform is initiated, Proposition \ref{Prop:ID}
 identifies scenarios where careerist agents exert higher implementation efforts under \textit{Full-Transparency} than under \textit{Semi-Transparency}. Therefore, the conventional wisdom on welfare comparison may not apply. Our results emphasize the need to consider implementation when assessing the welfare implication of transparency.

\subsection{Policy implications}
\subsubsection{Institution design}
Let's return to ``face innovation'' in China, where local officials pursue innovation for its mere appearance. Recent studies are debating the criteria for cadre promotions.\footnote{For instance, \cite{wang2021policy} argue that reform outcomes, rather than decisions, influence local officials' promotions.} Despite this, there's a widespread perception that face innovation is primarily a career advancement tool for policymakers. Local officials engage in this anticipating the unlikely negative career impacts of their policies. This aligns with our model's \textit{Non-Transparency} setting, where the principal lacks crucial policymaking information. Consequently, there is a pressing need for stricter regulations.

Ideally, a performance-based evaluation system that accounts for the consequences of policies would address face innovation. However, local officials' typically short tenures, usually less than three years, imply they frequently move before the effects of their policies are fully realized (e.g., \citet{teets2020evolution}).  As a result, evaluators may be compelled to base their decisions on immediate policy choices and initial implementations, rather than on long-term outcomes.  


Teets proposes a solution: link promotions to long-term results even after an official has transferred.\footnote{See \citet[p.103]{teets2015politics}.} The intent is to bind evaluators to a promotion rule contingent on both efforts and outcomes. Specifically, Teets describes an ideal system as follows:

\begin{quote}
``Redefining the `innovation' category on cadre evaluations to emphasize efforts to ensure sustainability
and create a learning system with long-term goals... This redefined innovation category would be measured by indicators
showing efforts at the institutionalization of innovations, such as empowering agencies to implement these policies and funding this implementation, and would also tie this innovation to its long-term performance.''
\end{quote}
We agree with \cite{teets2015politics}  that increasing transparency in policymaking is essential to curbing face innovation. However, we are skeptical that the proposed solution has addressed the commitment issue. Specifically, a political organization may struggle to penalize an agent who has demonstrated loyalty and competence by making the right decisions and implementing policies effectively, yet faces policy failure due to unforeseeable circumstances. Our findings suggest that no principal can credibly contract on both implementation efforts and outcomes. Solutions such as \cite{teets2015politics} may inadvertently encourage policymakers to behave in a complaisant or ``timid'' manner, detracting from the pursuit of successful reforms. 

Indeed, Teets's solutions  hold local officials accountable for both their decisions and implementations, as demonstrated in Proposition \ref{Prop:FT}. They encourage more responsible governance and can reduce face innovation. Yet, there are instances where a performance-based rule is more effective. Such a rule is credible only if the principal backs away from full transparency. If the central government wishes to apply a \cite{teets2015politics} rule to hold officials accountable for policy consequences (even after their positions have been transferred), it must be combined with local autonomy to protect implementation details from undue scrutiny. Otherwise, fearing inspections to reveal ``unsatisfactory progress,'' local officials are prone to act timidly. The recent \textit{System of Requesting Instructions and Submitting Reports} illustrates this: mandating local officials to report upward before acting ensures compliance, but it may also discourage innovative decisions \citep{chen2017u}.

\subsubsection{Motivating by accountability}\label{sec:AoN}
So far, we have examined the policymaking behaviors of careerist agents, showing that they can be motivated to pursue successful reforms under \textit{Semi-Transparency} and to diligently implement reforms under \textit{Full-Transparency}.

We now consider a situation where the principal delegates reform policymaking to a ``nonaccountable official,'' an agent who does not fear job loss  \citep{maskin2004politician}. Examples include  tenured judges, who are not at risk of losing their jobs over unpopular rulings, and ``lame duck'' politicians who, nearing the end of their terms, have diminished incentives to cater to public opinion. These officials are modeled as agents with no value in office-holding ($R=0$). \cite{maskin2004politician} note that these nonaccountable officials are often more motivated to gather costly information about the optimal policy choice. However, it is unclear  whether they are more motivated than accountable officials to choose and implement these decisions. 

We compare whether the principal necessarily benefits from delegating to an accountable official or a nonacountable one, within the context of multidimensional policymaking. In Lemma \ref{lemma:bad incentive} of the Appendix, we show that all types of nonaccountable agents will opt for the status quo, leading to a sure policyoutcome and no sorting for the principal. Conversely,
delegating to accountable agents provides the principal with some degree of sorting. If the principal can employ the right kind of transparency to motivate accountable officials, then delegating to them would be clearly preferable over nonaccountable ones. We further explore this comparison under the \textit{Semi-Transparency} information structure below.

\begin{Proposition}[Accountability or Non-Accountability]\label{Prop:AoN}
When the agent can be motivated $(\lambda R \geq 1)$, then reform decisions should  be allocated to accountable officials.
\end{Proposition}

Our analysis expands on the work of \citet{maskin2004politician} by further exploring the effects of motivation. We highlight that accountability is crucial for motivating implementation. Officials unaccountable for policy outcomes often  implement reforms at socially suboptimal levels due to the private costs of efforts. Conversely, accountable officials invest more efforts in implementation, as they are motivated to signal their competence through successful outcomes.

\subsubsection{Centralization}
Our paper builds on the key assumption that the agent is responsible for both deciding on and implementing reforms. In practice, however, the less informed principal may sometimes take back the authority to make policy choices, assigning only the implementation role to the agent. For instance, it is conceivable that Cummings directed department aides to support Brexit and enforced implementation quality through a system of rewards and penalties. 
For instance, it is conceivable that China's central government directed local officials to implement centrally crafted policies, with a strict monitoring system to ensure compliance. 
This kind of ``centralization'' reflects the principal's focus on control, even at the cost of sacrificing the agent's informational expertise—a longstanding theme in the delegation literature (e.g., \cite{holmstrom1980theory,bendor2001theories}).

We model the costs and benefits of this ``centralization'' by extending the baseline model. The core assumptions remain intact: the agent observes the nature of the reform while the principal does not. Reform success depends on both a ``good choice'' and ``good implementation.'' However, in this extended model, the less informed principal must decide between implementing a reform and maintaining the status quo. To make the problem interesting, we assume the principal's default policy choice is reform.\footnote{The scenario where the principal chooses between maintaining the status quo and delegating authority is less compelling. If the principal mandates the status quo, they receive the same total payoff as when delegating to a non-accountable agent—a situation already explored in Part \ref{sec:AoN}. Therefore, we do not delve into this case further.} Moreover, we assume the principal can enforce the agent to exert a minimum level of implementation effort, denoted as $\underline{e} \in (0,1]$. This parameter $\underline{e}$ reflects the principal's enforcement power. When $\underline{e} = 1$, the agent acts as the perfect implementer.

\begin{Proposition}\label{Centralization}
Suppose an uninformed principal instructs the agent to implement a reform. Such a decision is welfare-reducing unless all conditions are met: 1) the agent is unlikely to be congruent ($\pi$ is sufficiently small), 2)  a reform is likely good ($\phi$ is high), 3) the principal possesses the capability to enforce top-notch reform implementation ($\underline{e}$ is large), and 4) the political payoff is relative unimportant ($\alpha$ is low). 
\end{Proposition}

Compared with delegation, centralization ensures that the principal’s preferred policy is implemented. However, centralization can compromise the quality of policymaking. Beyond the traditional view that centralization prevents the principal from utilizing the agent’s expertise, it also limits the agent’s ability to signal their type through hard work. Worse still, the agent may shirk when enforcement power is weak during implementation. In such cases, there is a significant risk that the agent will exert less than optimal effort, which undermines effective policymaking.

In contrast, delegation with moderate oversight tends to address these challenges more effectively. Under \textit{Semi-Transparency}, the agent typically performs well: he chooses policies responsively to the reform's nature and implements them diligently when necessary. A lingering concern is the possibility of a misaligned agent who is not fully committed to the reform.  However, if most agents share the principal's preferences,  or if the few misaligned agents are still generally diligent, this concern becomes minor. In such cases, delegation serves the principal's interests.

\section{Conclusion}
In this paper, we examine the policymaking behaviors of careerist agents in reform decisions across three common information structures: \textit{Non-Transparency} (where only the reform decision is observable), \textit{Semi-Transparency} (where both the reform decision and policy outcomes are observable), and \textit{Full-Transparency} (where the reform decision, implementation details, and policy outcomes are observable). Unlike previous studies that typically model one-dimensional policymaking, we focus on scenarios where the agent is responsible for both selecting and implementing policies. In such a complex policymaking environment,  the right kind of transparency  is not merely a binary choice between ``transparency on action'' and ``transparency on consequences.''

Through equilibrium analysis, we show that a careerist agent is incentivized to pander to reform decisions when  career concerns are sufficiently strong.  Under such conditions, \textit{Non-Transparency} yields the most undesirable policy outcomes, as the agent is not held accountable for the implementation or its consequences. In contrast, \textit{Full-Transparency} increases the agent's accountability by expanding the range of observable actions and their consequences. However, the principal's electoral decisions lean heavily on observed actions than their actual outcomes. Thus, the principal cannot motivate the agent through performance-based reward schemes. Such rules become plausible under \textit{Semi-Transparency}, where careerist agents' only means of securing political rewards is to signal competence through successful reforms. Along the way, we compare the principal's payoffs across all information structures and identify the principal-optimal one.

Our results shed new light on classic debates in policymaking. Regarding whether decision-making authorities should be delegated or centralized, we show that delegation is more favorable when the policy gains  from  informed policymaking outweigh  the policy loss from losing control over misaligned agents. This finding suggests that the core principles of delegation theory are robust to the introduction of career concerns. Regarding whether technical decisions should be allocated to accountable or nonaccountable decision makers,  we highlight an additional benefit of accountability: it motivates implementation. By exercising a moderate level of control, the principal can motivate the agent to invest significant effort in the implementation process.

A key takeaway from our findings is that hierarchical organizations often benefit from relaxing control over lower-tier agents. In politics, writing explicit incentive contracts to motivate agents can be difficult. However, our results suggest that a principal (e.g, the central government) can effectively commit to performance-based incentives by granting greater decision-making autonomy to agents (e.g., local politicians), such as  maintaining an arm's length relationship with the agent. Stricter control and monitoring, on the other hand, can be counterproductive.

\clearpage

\renewcommand
\refname{Reference}
	\bibliographystyle{plainnat}
\bibliography{new.bib}

\newpage
\begin{appendices}

 	\setcounter{page}{1}

\section{Omitted Proofs}\label{Gen}

\subsection{Preliminary}\label{Pre}
\subsubsection{Incentives to implement}
We begin by describing the agent's incentive to implement a reform subject to different retention incentives.
\begin{lemma}\label{lemma:bad incentive}
Suppose the principal's retention does not vary with the outcome of a reform. Then,
\begin{enumerate}
    \item a congruent agent who chooses to reform will implement it with the effort $\lambda\mu $, where $\mu$ indicates whether $\omega=G$ ($\mu=1$) or not ($\mu=0$). Furthermore, his policy payoff from the reform is lower than that  from maintaining the status quo. 
    \item a noncongruent agent who chooses to reform will implement it with the effort $0$. Furthermore, his policy payoff from maintaining the status quo strictly exceeds that from  the reform.
    \end{enumerate}
\end{lemma}
\begin{proof}
The strategic behavior of the noncongruent agent is straightforward. Below we focus on the congruent agent's incentives. Conditional on initiating a reform, his objective is
\begin{align*}
    \max_e (2\mu e-1)v-\frac{e^2}{\lambda}.
\end{align*}
The agent's optimal effort is $\lambda \mu v$,  inducing payoffs from a reform $\lambda\mu^2 v^2-v$. Since $1\geq \lambda (v+R/2)>\lambda v$ by the assumptions in Section \ref{IA}, $\lambda\mu^2 v^2-v\leq v(\lambda v-1)< 0$. 
The results thus follow.
\end{proof}
\begin{lemma}\label{lemma:good incentive}
Suppose the principal retains if she observes a successful reform, and replaces otherwise. Then 
\begin{enumerate}
    \item a congruent agent who chooses to reform will implement it with the effort $\lambda\mu (v+\frac{R}{2}) $, where where $\mu$ indicates whether $\omega=G$ ($\mu=1$) or not ($\mu=0$). Furthermore, his total payoff from the reform exceeds that from maintaining the status quo if and only if $\lambda \mu^2 \geq \frac{v}{(v+R/2)^2}$.
    \item a noncongruent agent who chooses to reform will implement it with the effort $\lambda\mu \frac{R}{2} $, where $\mu$ indicates whether $\omega=G$ ($\mu=1$) or not ($\mu=0$). Furthermore, his total payoff from the reform exceeds that from maintaining the status quo if and only if $\lambda \mu^2 \geq \frac{4v}{R^2}$.
    \end{enumerate}
\end{lemma}
\begin{proof}
Rewrite the agent's objective function as 
\begin{align*}
  & \text{Congruent}\qquad \max_e \mu e(2v+R)-v-\frac{e^2}{\lambda}\\
   & \text{Nonongruent}\qquad \max_e \mu eR-v-\frac{e^2}{\lambda}.
\end{align*}
The congruent agent's optimal effort is $\lambda \mu (v+R/2)$,  inducing a payoff from a reform $\lambda\mu^2 (v+R/2)^2-v$. The noncongruent agent's optimal effort is $\lambda \mu R/2$.
The results thus follow.\end{proof}

\subsubsection{Off-path beliefs}
Now, we assign off-path beliefs using the divinity criterion. 
\begin{Definition}
Let $I\in \mathcal{I}$ be an arbitrary event observed by the principal. $I$ is {\bf neutral} news about the agent's congruence if $P(t=g|I)=\pi$, is {\bf good}  news if $P(t=c|I)>\pi$, and is {\bf bad} news if $P(t=c|I)<\pi$.
\end{Definition}
\begin{lemma}\label{lemma:belief}
A PBE surviving the universal divinity refinement assigns the following off-path beliefs:
\begin{enumerate}
\item if the status quo is off-path, then the
principal believes that whoever chooses it is congruent with probability $\frac{\pi(1-\phi)}{1-\pi\phi}$, which is smaller than $\pi$. 
\item if the reform is off-path, then the
principal believes that whoever chooses it is congruent with probability $1$.
\end{enumerate}
\end{lemma}
\begin{proof} 
Following \cite{fudenberg1991game}, we let $D((t,\omega), \mathcal{M}, x')$, $D^0((t,\omega), \mathcal{M}, x')$ be the set of the principal's mixed-strategy best response -- the set of principal's retention probability -- to the agent's action $x'$ and belief concentrated on $\mathcal{M}\subset \{C,N\}\times \{G,B\}$ that makes a type $(t,\omega)$ agent strictly benefits (indifferent) by taking $x'$ relative to his equilibrium action. Here $t\in \{C,N\}$ is the agent's congruence type while $\omega \in \{G,B\}$ is realization of the state. The divinity condition says that fixing some off path action $x'$, if for some $\omega \in \{G,B\}$ and $t\in\{C, N\}$ we have $D((t,\omega), \mathcal{M}, x')\cup D^0((t,\omega), \mathcal{M}, x')\subset \bigcup_{(t',s')\neq (t,x)}D((t',s'), \mathcal{M}, x')$, then we can assign probability $0$ to the event that the deviation $x'$ comes from type $(t,\omega)$. Informally, this condition says that the deviation $x'$ is unlikely to come from a type $(t,\omega)$ agent, so we rule out it first. After that, we iterate this process until it ends. 

Among four types of the agent, $(N,G)$, $(N,B)$, and $(C, B)$ share the same policy preferences. This is because they derive a policy payoff of $-v$ by initiating a reform net of the implementation cost, and $0$ by maintaining the status quo. Since a type $(C,G)$ agent is more motivated to reform than types $(C,B),(N,G), (N,B)$, establishing the lemma amounts to showing:
\begin{itemize}
  \item If the status quo is off path, the  type $(C,G)$ is ruled out if the principal observes $(q,0)$.
    \item If the reform is off-path, all types other than $(C,G)$ are ruled out if the principal observes $(r,e)$ for some $e\in [0,1]$,
\end{itemize} 

For expository convenience, we notate $x'=r$ as an agent's action if the reform is off the equilibrium path. Similarly, we notate $x'=q$ if the status quo is off path. 

 Let $\underline{p}^{(C,\omega)} (\mathcal{M},x')$ be the retention probability with which a type $(C,\omega)$ agent will receive the same payoff as his equilibrium payoff after he deviates to action $x'$. Similarly, we define $\underline{p}^{(N,\omega)} (\mathcal{M},x')$. Then $D((C, \omega), \mathcal{M}, x')=\{p:p> \underline{p}^{(C,\omega)}(\mathcal{M},x')\}$ and $D((N,\omega), \mathcal{M}, x')=\{p:p>\underline{p}^{(N,\omega)}(\mathcal{M},x')\}$ where $p$ is the retention probability. Since $(N,G)$, $(N,B)$, and $(C, B)$ share the same policy preferences, $\underline{p}^{(C,B)}(\mathcal{M},x')=\underline{p}^{(N,G)}(\mathcal{M},x')=\underline{p}^{(N,B)}(\mathcal{M},x')$ for all $x'\in \{r,q\}$. To prove the claim, it is sufficient to verify whether $\underline{p}^{(C,G)}(\mathcal{M},q)>\underline{p}^{(C,B)}(\mathcal{M},q)$ and  
 $\underline{p}^{(C,G)}(\mathcal{M},r)<\underline{p}^{(C,B)}(\mathcal{M},r)$. Since we focus on equilibria where the principal uses a pure strategy of retention, along the path the equilibrium retention probability $\sigma^*\in \{0,1\}$.

{\bf When the status quo is off-path:} In equilibrium, every type of the agent panders to reform, and the principal retains on path. Let $e^*_{(t,\omega)}$ be the equilibrium effort of a type $(t,\omega)$ agent. We will describe the structure of the equilibrium first.  

Case I: If implementation efforts are observable, then we claim that there exists a constant $e^*\in [0,1]$ such that $e^*_{(t,\omega)}=e^*$ for all types $(t,\omega)$. To show this, let $E^*$ be the set of implementation efforts that would lead to retention. $E^*\neq \emptyset$ since all types pander to reform and are retained. Then all types but a type $(C,G)$ agent will choose the action $(r, \underline{e})$, where $\underline{e}$ is the minimal element of  $E^*$. Now, if a type $(C,G)$ agent chooses the implementation effort $e^*\in E^*$ but $e^*>\underline{e}$, then $(r, \underline{e})$ becomes bad news about the agent's congruence, thus leading to replacement. Contradiction. Thus, it must be that 
$e^*=\underline{e}$.  Hence, any equilibrium where the $q$ is off path must take the form: all types of the agent reforms with the same implementation effort, which we denote by $e^*$.

Case II: If efforts are not observable then $e^*_{(N,\omega)}=e^*_{(C,B)}=0$ and $e^*_{(C,G)}\in \arg\max (2e-1)v-\frac{e^2}{\lambda}$. This follows from the requirement of sequential rationality in any PBE. 

Now we are in a good position to prove Part I of the lemma. The definition of $\underline{p}$ requires that if efforts are observable, $\underline{p}^{(C,G)}(\mathcal{M},q)\cdot R= R+(2 e^*-1)v-\frac{{e^*}^2}{\lambda}$ and $\underline{p}^{(C,B)}(\mathcal{M},q)\cdot R= R-v-\frac{{e^*}^2}{\lambda}$. If efforts are unobservable then $\underline{p}^{(C,G)}(\mathcal{M},q)\cdot R= R+\max_e\{(2 e-1)v-\frac{{e}^2}{\lambda}\}$ and $\underline{p}^{(C,B)}(\mathcal{M},q)\cdot R= R-v-0$. One can conclude that in both cases $1>\underline{p}^{(C,G)}(\mathcal{M},q)>\underline{p}^{C,B}(\mathcal{M},q)$; the first inequality holds because $(2 e^*-1)v-\frac{{e^*}^2}{\lambda}\leq \max_e\{(2 e-1)v-\frac{{e}^2}{\lambda}\}=\lambda v^2-v< 0$. In other words, types $(N,G)$, $(N,B)$, and $(C, B)$ have more to gain by deviating to the status quo policy than the type $(C,G)$. This implies that  we can rule out the type $(C,G)$ after observing a deviation to the status quo, and the iteration process stops. We thus compute that $P(t=C|x=q)=\frac{\pi(1-\phi)}{1-\pi \phi}<\pi$. 

{\bf When the reform is off-path:} By the same logic, the type $(C,G)$ has more to gain from such a deviation than other types. We briefly sketch the steps here: fix a pooling equilibrium at $x=q$, and consider the deviation to $x=r$ with any nonnegative implementation effort $e'$. Since the type $(C,G)$ always obtains strictly more policy payoff than other types from such a deviation, $\underline{p}^{(C,G)}(\mathcal{M},r)<\underline{p}^{C,B}(\mathcal{M},r)=\underline{p}^{N,\omega}(\mathcal{M},r)$. $\underline{p}^{(C,G)}(\mathcal{M},r) R+(2e'-1)v-\frac{e'^2}{\lambda}= R\cdot \sigma^*$, where $\sigma^*$ indicates whether the principal retains along the path ($\sigma^*=1$) or not ($\sigma^*=0$). By Lemma \ref{lemma:bad incentive}, LHS $<\underline{p}^{(C,G)}(\mathcal{M},r)\cdot R\leq R$. Thus, if the principal retains along the equilibrium path ($\sigma^*=1$), no type can profitably deviate to the reform. In this case, the divinity criterion does not apply, so we assign probability one to the type $(C,G)$ following a reform decision. Otherwise, $\underline{p}^{(C,G)}(\mathcal{M},r)\in (0,1)$, and  the principal assigns probability one to the type $(C,G)$ following a reform decision by the divinity criterion. \end{proof}

\subsubsection{Inaction equilibrium}
\begin{lemma}[Inaction equilibrium]\label{Eq: inaction}
An equilibrium exists for any environment $J\in \{NT, ST, FT\}$: all types of the agent choose the status quo, and are always retained by the principal. 
\end{lemma}
\begin{proof}
We verify whether the described strategies satisfy the equilibrium conditions. When all types choose the status quo along the path, the principal's posterior belief is $P(t=C|x=q)=\pi$, which justifies a retention decision. Off the path, Lemma \ref{lemma:belief} specifies that whoever reforms is congruent, which justifies a retention decision. Thus, the principal's retention strategy is sequentially rational. Since the agent's retention does not hinge on the policy choice, he will opt for his optimal policy action. By Lemma \ref{lemma:bad incentive}, the agent cannot benefit from initiaing a reform because $\lambda<1/v$. Thus, all types of the agent will not deviate from the equilibrium strategy.  
\end{proof}

\subsection{Proof of Proposition \ref{Prop:NT}}
\textbf{Existence.} We specify the principal's sequentially rational retention strategy. When every type chooses to reform, the principal believes
that whoever chooses the reform is congruent with probability $\pi$, and whoever deviates to the status quo is congruent with probability $\frac{\pi(1-\phi)}{1-\pi \phi}<\pi$ by Lemma  \ref{lemma:belief}.  Thus, it is sequentially rational for the principal to retain the agent when she observes a reform decision ($x=r$), and replace when she observes the status quo ($x=q$). 

Now we check whether the strategies described in the Proposition satisfy the equilibrium conditions. Along the path, the principal retains whoever reforms; off the path, the principal replaces whoever deviates to the status quo. Given this retention strategy, the agent who chooses to reform will receive a payoff no less than $R-v$, which is larger than the payoff from deviating to the status quo, $0$.  Thus, all types of the agent will adhere to the equilibrium strategies.

\textbf{Uniqueness.} We rule out other equilibrium possibilities.  Lemma \ref{Eq: inaction} identifies an inaction equilibrium where only the status quo is chosen along the equilibrium path; Proposition \ref{Prop:NT} identifies an equilibrium where only the reform is chosen along the equilibrium path. Thus, the remaining possibilities are that both policies ($x=r$ and $x=q$) are chosen along the equilibrium path. 

We do so by examining the principal's retention strategies exhaustively. [a] Suppose the principal retains after observing both policies; both policies occur along the path. By Lemma \ref{lemma:bad incentive}, all types' sequentially rational strategy is to choose the status quo, contradicting that  ``both policies occur along the path.'' [b] Suppose the principal retains after observing a reform and replaces after observing the status quo; both policies occur along the path. Then whoever chooses the status quo along the path has a profitable deviation to choosing the reform with implementation effort $0$. Contradiction. [c] Suppose the principal retains after observing the status quo and replaces after observing a reform; both policies occur along the path. Then optimality  requires all types to choose the status quo for sure, contradicting that ``both policies occur along the path.'' [d] Suppose the principal replaces after observing both policies; both policies occur along the path. Then by Lemma \ref{lemma:bad incentive}, all types choose the status quo for sure, contradicting that ``both policies occur along the path.'' So we are done.

\subsection{Proof of Proposition \ref{Prop:ST}}
\textbf{Existence.} We first check whether the strategies described in Proposition \ref{Prop:ST} satisfy the equilibrium conditions. Fix the agent's action. When the principal observes a successful reform, her posterior belief is that $P(t=C|r, S)=\frac{\pi \lambda (v+R/2) }{\pi \lambda (v+R/2)+(1-\pi) \lambda R/2}>\pi$, which justifies a retention decision.  When the principal observes a failed reform, her posterior belief is that $P(t=C|r, F)=\frac{\pi (1-\lambda (v+R/2)) }{\pi (1-\lambda (v+R/2))+(1-\pi) (1-\lambda R/2)}<\pi$, which justifies a replacement. When the principal observes a status quo, her posterior belief is $P(t=C|q)=\pi$, which justifies a replacement. Thus, the principal's retention strategy is sequentially rational. Now we fix the retention strategy. By Lemma \ref{lemma:good incentive}, the efforts described in Proposition \ref{Prop:ST} are optimal conditional on a reform. The type $(N,G)$ agent's equilibrium strategy brings him a payoff of $\lambda (R/2)^2-v$, which is larger than the payoff from deviating to maintaining the status quo. Likewise, the type $(C,G)$ agent's equilibrium strategy brings him a payoff of $\lambda (v+R/2)^2-v$, which is larger than the payoff from deviating to maintaining the status quo. Furthermore, types $(C,B)$, $(N,B)$ will not deviate to initiating a reform, as the reform will surely fail and the replacement ensues. Thus, the equilibrium conditions are met.

\textbf{Uniqueness.} We rule out other equilibrium possibilities other than inaction (i.e. all types choose the status quo).  

[a] Only one type of $(t,\omega)\in \{C,N\}\times \{G,B\}$ reforms with a strict positive probability along the path. Given the fact that the type $(C,G)$ receives a strictly higher policy payoff from a reform than other types (who share the same policy payoff), such a strategy is plausible only if $(C,G)$ reforms while all other types choose the status quo; otherwise,  $(C,G)$ has a strict incentive to deviate to initiating a reform. Then, maintaining the status quo $\{x=q\}$ becomes bad news about the agent's congruence, while initiating a reform $\{x=r\}$
becomes good news. This means that the principal should retain the agent after observing a reform and replace this agent after observing the status quo. Given this retention strategy, the type $(C,G)$ agent will choose the effort $\lambda v$ by Lemma \ref{lemma:bad incentive}. Since $\lambda v<1$, the reform will fail with a positive probability along the equilibrium path. This creates incentives for the type $(C,B)$ (and other types other than $(C,G)$) to deviate to choosing the reform with implementation effort $0$, as the principal will retain after observing a reform even if it fails. The gain from such a deviation is $R-v>0$, which exceeds the payoff from maintaining the status quo.  

[b] Two types from
$(t,\omega)\in \{C,N\}\times \{G,B\}$ reform with a strict positive probability along the path.  We have constructed an equilibrium where $(C,G)$ and $(N,G)$ reform while  $(C,B)$ and $(N,B)$ choose the status quo. Notably, if at least one type of the form $(\cdot,B)$ chooses to reform, then a failed reform must be rewarded retention. The possibilities are: 

If types $(C,G)$ and $(C,B)$ reform while $(N, G)$ and $(N,B)$ choose the status quo, the principal retains after observing a reform and replaces after observing the status quo. Then, since all reforms decisions are carried out by the congruent agent and failure occurs along the path, types $(N,\cdot)$ can deviate to choosing a reform with implementation effort $0$ and obtain a payoff of $R-v$, which is larger than their equilibrium payoff $0$.

If types $(C,G)$ and $(N,B)$ reform while $(N, G)$ and $(C,B)$ choose the status quo, the principal retains after observing a \textit{failed} reform on path and replaces after observing the status quo. Otherwise, $(N,B)$ could profitably deviate to the status quo. However, given this retention incentive, both types $(C,B)$ and $(N,G)$ can profitably deviate to choosing a reform with implementation effort $0$. Similar logic applies to the situations where 
(1) types $(N,G)$ and $(C,B)$ reform while $(C, G)$ and $(N,B)$ choose the status quo, and (2) types $(C,B)$ and $(N,B)$ reform while $(C, G)$ and $(N,G)$ choose the status quo. (3) types $(N,G)$ and $(N,B)$ reform while $(C, G)$ and $(C,B)$ choose the status quo. In all three cases, at least one type of the form $(\cdot, B)$ reforms along the path, meaning that a failed reform is rewarded. This creates incentive for at least one other type to reform with implementation effort $0$.

[c] Three types from
$(t,\omega)\in \{C,N\}\times \{G,B\}$ reform with a strict positive probability along the path. We deduce what is the only type that refrains from the reform. Firstly, the type cannot be $(C,G)$, because such an agent receives strictly higher payoffs than other types. Secondly, the type cannot be $(C,B)$. In this case, the type $(N, B)$ reform with probabilities, meaning that  along the path (1) a failed reform is  rewarded retention, (2) the status quo is punished by replacement. In this case, the type $(C,B)$ should also reform.  For the same reason, thirdly, the type cannot be either $(N,B)$ or $(N,G)$. Thus we have exhausted all possibilities. 

[d] All types reform with strict positive probabilities.  Then, the principal must retain after observing a failed reform, and replace after observing the status quo (if any one of two conditions do not hold, types of the form $(\cdot, B)$ can profitably deviate to the status quo). (1) If the principal retains after observing a successful reform, 
then the retention decision does not vary with reform outcomes. Since a reform decision is valued at $R$ which exceeds the stake of a failed reform $v$, all types will pander to  reform with probability one. Moreover,  by Lemma \ref{lemma:bad incentive}, only $(C, G)$ reforms with positive implementation efforts, leading to posterior beliefs $P(t=C|r, S)>\pi $ and $P(t=C|r, F)<\pi$. Thus, the principal's sequentially rational retention strategy should be to replace an agent who unsuccessfully reforms, contradiction. (2) If the principal replaces after observing a successful reform, then along the equilibrium path, all types choose to reform with implementation effort $0$. In equilibrium, successful reforms are off the equilibrium path, and each suffers from a failed reform. However, repeating the steps in Lemma \ref{lemma:belief}, one concludes that divinity criterion assigns belief $P(t=C|r, S)=1$. This is because the type $(C,G)$ is the only type that can both (1) successfully reform (2) benefit from deviating to $(r, e)$ with $e>0$. In fact, $(r,\lambda v)$ is a profitable deviation for the type $(C,G)$. Thus, the conjecture profile does not survive the divinity criterion. 

\subsection{Proof of Proposition \ref{Prop:FT}}
{\bf Existence.} As we have illustrated, types $(C,B)$, $(N,G)$, and $(N,B)$ share the same policy preference: they prefer the status quo to the reform. The type $(C,G)$ receives a strictly higher reform payoff than these types. This alludes to the possibility of a separating equilibrium: the type $(C,G)$ may reform with an implementation effort such that all other types find it too costly to mimic;  the principal retains after observing a reform decision with a ``satisfying'' level of implementation effort, and replaces after observing either a poorly implemented reform or the status quo.

The least costly effort $e_S$ that achieves separation is determined by the equation $R-v=\frac{e^2_S}{\lambda}$, or $e_S=\sqrt{\lambda (R-v)}$. Further, note that the reform payoff of a type $(C,G)$ agent is single-peaked at the $\lambda v$. Thus, in such a separating equilibrium, a type $(C,G)$ who reforms will exert effort $e^*=\max\{\sqrt{\lambda(R-v)}, \lambda v\}$. Other types choose the status quo. Along the path, the principal retains after observing $(r,e^*)$ and replaces after observing the status quo. Now, we use the universal divinity refinement to specify the off-path beliefs, and accordingly complete the description of the equilibrium strategies. 

Let's first consider whether any type may benefit from a deviation to choosing a reform with implementation effort $e'\neq e^*$.  It is straightforward to verify that no type may benefit if $e'>e^*$, so we focus on the cases where $e'\in [0,e^*)$. 

(1) $\lambda v^2\geq R-v$. In this case, $e^*=\lambda v$ so we will verify whether any type may gain by deviating to the action $(r,e')$ where $e'<\lambda v$. The type $(C,G)$ will not, because by Lemma \ref{lemma:bad incentive}, the implementation effort maximizes his policy payoff conditional on a reform decision. Thus, the types that may gain from such a deviation are types $(C,B)$, $(N,G)$, and $(N,B)$. This means the divinity criterion will assign the following off-path belief: for $e'\in [0,e^*), P(t=C|(r,e'))=\frac{(1-\phi) \pi}{1-\phi \pi}<\pi$, so the principal should replace the agent after observing it. 

(2)  $\lambda v^2< R-v$. In this case, $e^*=\sqrt{\lambda (R-v)}$ so we will verify whether any type may gain by deviating to the action $(r,e')$ where $e'<\sqrt{\lambda (R-v)}$. 

We again use the notation $\underline{p}^{(t,\omega)}$ to represent the retention probability that make a type $(t,\omega)$ agent indifferent between the equilibrium action and the deviation $(r,e')$. As before, since types $(C,B)$, $(N,G)$, and $(N,B)$ share the same policy preference, $\underline{p}^{(C,B)}=\underline{p}^{(N,G)}=\underline{p}^{(N,B)}$. We aim to show that  $\underline{p}^{(C,B)}\geq \underline{p}^{(C,G)}$ i.e., a type$(C,B)$ agent benefits more from this deviation and thus can tolerate more retention loss.  By definition, $\underline{p}^{(C,B)}\cdot R-v-\frac{{e'}^2}{\lambda}=0$; $\underline{p}^{(C,G)}\cdot R+ (2e'-1)v-\frac{{e'}^2}{\lambda}= R+(2e^*-1)v-\frac{{e^*}^2}{\lambda}=2e^*v$, which simplifies to $\underline{p}^{(C,G)}\cdot R+ 2(e'-e^*)-v-\frac{{e'}^2}{\lambda}= 0$. 
Straightforward comparison suggests that $\underline{p}^{(C,B)}<\underline{p}^{(C,G)}$, meaning that the types $(C,B)$ and $(N,\omega)$ have more to gain from this deviation; furthermore, $\underline{p}^{(C,B)}\in(0,1)$.  This is because the type $(C,B)$ agent has less stake in reform, and he does not suffer as much as the type $(C,G)$ in reducing efforts. The principal assigns the following off-path belief: for $e'\in [0,e^*), P(t=C|(r,e'))=\frac{(1-\phi) \pi}{1-\phi \pi}<\pi$. 

Thus, we have characterized the least costly separating equilibrium (or the Riley outcome) that survives the universal divinity refinement: a type $(C,G)$  will reform with implementation effort $e^*=\max\{\sqrt{\lambda(R-v)}, \lambda v\}$. Other types choose the status quo. Along the path, the principal retains after observing $(r,e^*)$ and replaces after observing the status quo. The beliefs are $P(t=C|(r,e^*))=1$ and $P(t=C|q)=\frac{(1-\phi) \pi}{1-\phi \pi}$. Off the path, the divinity criterion dictates that for $e'\in [0,e^*), P(t=C|(r,e'))=\frac{(1-\phi) \pi}{1-\phi \pi}<\pi$; whenever an action $(r,e')$ is observed, the principal should replace the agent. When an action $(r,e'')$ with $e''>e^*$ is observed, the divinity criterion does not apply, so we assign the belief  $P(t=C|(r,e'))=1$ and the principal should retain the agent after seeing it.

{\bf Uniqueness.} First, we consider other equilibria where the type $(C,G)$ separates. Recall that $\sqrt{\lambda (R-v)}$ is the minimal effort to deter other types from mimicking the type $(C,G)$. 
Were there any other separating possibility, it must be that the type $(C,G)$ reforms with an implementation effort higher than this, while other types maintain the status quo. Thus, we consider whether the following strategies may constitute part of an equilibrium: the type $(C,G)$ reforms with implementation effort 
$\max\{\Tilde{e}, \lambda v\}$ where $\Tilde{e}>\sqrt{\lambda(R-v)} $; other types maintain the status quo. Depending on the parameters there are two cases. (1) $\lambda v^2\geq R-v$. Then the only equilibrium possibility that differs from the one described in Proposition \ref{Prop:FT} is when $\Tilde{e}>\lambda v$. However, in this case, the type $(C,G)$ can profitably deviate to initiating a reform with implementation effort $e^*=\lambda v$. To see this, none of the types $(C,B)$, $(N,G)$, and $(N,B)$ can profitably deviate to this strategy $(r,\lambda v)$ even if it secures retention. Thus, the principal assigns the off-path belief $P(t=C|(r,\lambda v))=1$ using the divinity condition, and retains the agent after observing $(r,\lambda v)$. (2) $\lambda v^2< R-v$. Then we shall consider the possibility where $\Tilde{e}>e^*=\sqrt{\lambda (R-v)}$. For the same reasoning as in the previous paragraph, the type $(C,G)$ can profitably deviate to initiating a reform with implementation effort $e^*=\sqrt{\lambda (R-v)}$. Therefore, the separating equilibrium described in Proposition \ref{Prop:FT} is the only separating equilibrium that survives the divinity condition.

Next, we examine whether equilibrium possibilities exist where the type $(C,G)$ reforms, and  at least one type denoted $(\hat{t},\hat{\omega})\neq (C,G)$ reforms and implements with effort $\hat{e}\in [0,1]$. 

[a] Exactly one type reforms. (a1) If this type is $(C,B)$, the principal retains after observing a reform decision, and replaces after observing the status quo. However, since types $(N,\omega)$ share the same policy payoff with $(C,B)$, they have incentives to deviate from the status quo to the action $(r,\hat{e})$. (a2) If this type is $(N,G)$, then it must be that  the principal  replaces after observing the status quo. Otherwise, $(N,G)$ should deviate to the status quo. In this situation, the effort choice of $(N,G)$ must be the same with the effort of $(C,G)$, for otherwise the principal can perfectly identify the types. So the conjectured equilibrium has the following structure: both the types $(C,G)$  and $(N,G)$ choose to reform with the same implementation effort $\hat{e}$, while types $(C,B)$ and $(N,B)$ maintain the status quo; the principal replaces after observing the status quo. Now consider the principal's retention decision after a reform is initiated. If the principal retains after observing a failed reform, then $(C,B),(N,B)$ can profitably deviate to the action $(r,\hat{e})$, because in doing so they receive a weakly higher payoff than the type $(N,G)$. If the principal does not retain after observing a failed reform, then she must retain after observing a successful reform, for otherwise the type $(N,G)$ can profitably deviate to the status quo. If the principal retains after a successful reform while replaces after an unsuccessful reform, then either one of the types $(C,G)$ and $(N,G)$ does not optimize his implementation effort by Lemma \ref{lemma:good incentive}. Thus, we have ruled out this possibility. (a3) If this type is $(N,B)$, then the principal must retain after observing a failed reform. Otherwise, the type $(N,B)$ can profitably deviate to the status quo. However, this retention strategies incentivize the type $(C,B)$ to deviate to $(r,\hat{e})$. 

[b] Exactly two out of three types $(C,B)$,$(N,G)$, and $(N,B)$ reform. Then at least one agent who observes $\omega=B$ chooses to reform along the equilibrium path, and such a reform fails for sure. 
Sequential rationality implies that the principal's equilibrium retention strategy must be that she will retain after observing a failed reform; otherwise, the agent who observes $\omega=B$ should opt for the status quo. However, this creates incentive for the remaining type among the three to deviate to the action $(r,\hat{e})$. Contradiction. 

[c] All types reform. For the same reasoning as before, the principal should retain an agent even if his reform fails. (c1) If the principal does not retain an agent who successfully reform, the all types will best respond to this retention strategy by exerting zero effort along the path (i.e. choose the action ($r,0$)). In this situation, however, the type $(C,G)$ can deviate to the action $(r, \lambda v)$, which is off the path. The divinity criterion assigns belief $P(t=C|(r,\lambda v))=1$, which means that the principal's retention strategy is not sequentially rational. (c2) If the principal retain an agent who successfully reforms, then reform outcomes do not matter for retention. Let $E$ be the set such that the principal retains the agent after an action $(r,e)$ with $e\in E$ is observed. Since all types but $(C,G)$ incur a policy loss of $v$ by initiating a reform, it must be that all three types pool on the minimal implementation effort that secures retention, which we denote by $\hat{e}$. Moreover, the type $(C,G)$ must also choose implementation effort $\hat{e}$ along the path, for otherwise $(r,\hat{e})$ becomes bad news for retention. This alludes to the following equilibrium possibility: 

All the types choose the action $(r,\hat{e})$ along the equilibrium path, where $\hat{e}$ is to be determined. The principal retains after observing $(r,\hat{e})$, and replaces after observing the status quo.

We shall complete the equilibrium characterization using the divinity condition. There are a few steps:

Claim: In any pooling equilibrium that survives the divinity refinement, it must be that the agent chooses the action $(r,\hat{e})$ where  $\hat{e}\geq \lambda v$. 
\begin{proof}
Suppose not, which means that all types of the agent pool on the action $(r,\hat{e})$ where  $\hat{e}< \lambda v$. In this case, the type $(C,G)$ can profitably deviate to $(r, \lambda v)$. Repeating the steps as in the proof of Lemma \ref{lemma:belief}, we conclude that  the divine condition assigns belief $P(t=C|(r,\lambda v))=1$, so the type $(C,G)$ will be retained after this deviation. Since $\lambda v$ is the type $(C,G)$'s optimal implementation effort, he strictly benefits from this deviation.  This contradicts the equilibrium condition.  
\end{proof}

Now, fix an equilibrium where all the types choose the action $(r,\hat{e})$ with $\hat{e}\geq \lambda v$ along the equilibrium path, we use the divinity condition to pin down the off-path beliefs. Along the path, the principal retains everyone. We consider 
the case where a deviation to some $(r,e')$ where $e'\neq e^*$ occurs.  As before, we define $\underline{p}^{(t,\omega)}$ as the retention probability that a type $(t,\omega)$ agent will receive the same payoff as his equilibrium payoff after he deviates (the exact action is omitted). $\underline{p}^{(C, B)}=\underline{p}^{(N, B)}=\underline{p}^{(N, G)}$ since they share the same policy preference. By definition,
\begin{align*}
    R\underline{p}^{(C,G)}+(2 e'-1)v-\frac{(e')^2}{\lambda }&=R+(2 e^*-1)v-\frac{(e^*)^2}{\lambda}\\
            R\underline{p}^{(C,B)}-v-\frac{(e')^2}{\lambda }&=R-v-\frac{(e^*)^2}{\lambda}.
\end{align*}
By comparison, we deduce that $\underline{p}^{(C,G)}\geq \underline{p}^{(C,B)}\Leftrightarrow e^*\geq e'$. Informally, whenever an implementation effort lower (higher) than the equilibrium level occurs, the principal believes that it is more likely to come from the noncongruent (congruent) types.  Thus, she will replace whenever any deviation $(r,e')$ with $e'<e^*$ occurs, and retains whenever any deviation $(r,e'')$ with $e''>e^*$ occurs. 

Similarly,  repeat the above steps and modify the deviation to $q$, we have 
\begin{align*}
    R\underline{p}^{(C,G)}-v&=R+(2 e^*-1)v-\frac{(e^*)^2}{2\lambda}\\
            R\underline{p}^{(C,B)}-v&=R-v-\frac{(e^*)^2}{2\lambda}.
\end{align*}
Thus, the principal believes that whoever deviates to the status quo is likely noncongruent, and she will replace accordingly. 

Finally, we should restrict $\hat{e}$ such that the agent finds it optimal to choose $(r,\hat{e})$ where $\hat{e}\geq \lambda v$. Clearly, no type wants to deviate to initiating a reform with an implementation effort higher than $\hat{e}$. No type wants to deviate to initiating a reform with an implementation effort lower than $\hat{e}$. This is because doing so forgoes retention, and choosing the status quo is a better deviation (Lemma \ref{lemma:bad incentive}). Thus, it remains to specify conditions such that $(r,\hat{e})$ is preferred to the status quo for types $(C,B)$ and $ (N,\omega)$. Their equilibrium payoff is $R-v-\frac{\hat{e}^2}{\lambda}$, which is higher than the status quo payoff $0$ if and only if $\hat{e}\leq \sqrt{\lambda (R-v)}$. Thus, if $\lambda v\leq \sqrt{\lambda (R-v)}$, which amounts to $\lambda v^2\leq R-v$, the following pooling equilibrium survives the divinity refinement: 

{\it All types of the agent pool on the action $(r,e^*)$ with $e^*\in [ \lambda v, \sqrt{\lambda(R-v)}]$. The principal believes that $P(t=C|(r,e^*))=\pi$ and retains along the path. Off the path, the principal believes that $P(t=C|(r,\hat{e}))=1$ for $\hat{e}>e^*$ and retains, and $P(t=C|(r,\hat{e}))=P(t=C|q)=\frac{(1-\phi)\pi}{1-\pi \phi}$ for $\hat{e}<e^*$ and replaces. }

Thus we have completed the proof.

\subsection{Proof of Proposition \ref{Prop:ID}}

\begin{lemma}[Condition M]\label{CondM}
   If  Condition M holds, a type $(C,G)$ agent exerts more implementation effort under \textit{Full-Transparency} than under \textit{Semi-Transparency}.
\end{lemma}
\begin{proof}
    
 We will compare whether a type $(C,G)$ agent exerts more implementation effort under \textit{Semi-Transparency} or under \textit{Full-Transparency}. This amounts to comparing two values: $\lambda (v+\frac{R}{2})$ and $\max \{\lambda v,\sqrt{\lambda (R-v)}\}$, so we check  whether $\lambda (v+\frac{R}{2})^2\geq R-v$ holds.  Define $H(R)=\frac{\lambda}{4}R^2+(\lambda v-1)R+(\lambda v^2+v)$, so that 
 $\lambda (v+\frac{R}{2})^2\geq R-v$ if and only if $H(R)\geq 0$. 
$H(R)=0$ has real solutions if and only if $(\lambda v-1)^2\geq \lambda(\lambda v^2+v)$ or $3\lambda v\leq 1$. In this case, $H(R)$ has two solutions 
$\bar{R}(\lambda,v) =\frac{1-\lambda v+\sqrt{1-3\lambda v}}{\lambda/2}$ and $\underline{R}=\frac{ 1-\lambda v-\sqrt{1-3\lambda v}}{\lambda/2}$. Thus, if both $3\lambda v< 1$ and $R\in (\underline{R},\bar{R})$ hold, then  a type $(C,G)$ agent exert more effort under \textit{Full-Transparency} than \textit{Semi-Transparency}. Otherwise, \textit{Semi-Transparency} motivates better. 
\end{proof}
\begin{proof}[Proof of Proposition \ref{Prop:ID}]
(1) We first show that $W^{NT}<W^{ST}$. With a pairwise comparison, we can see that the principal receives a strictly higher policy payoff ($2 \pi \phi \lambda v^2+\phi \lambda Rv-\phi v>2 \pi \phi \lambda v^2-v$) and a strictly higher political payoff ($\pi[1+(1-\pi)\phi \lambda v ]>\pi$) from using \textit{Semi-Transparency} than using \textit{Non-Transparency}. We next show that $W^{NT}\leq W^{FT}$. When $\lambda v^2> R-v$, a semi-responsive equilibrium exists. The principal's expected payoff is $W^{FT}=   2\pi \phi \max \{\sqrt{\lambda (R-v)}, \lambda v\}v -\pi \phi v+\alpha \pi [1+(1-\pi)\phi]>2 \pi \phi \lambda v^2-v+\alpha \pi=W^{NT}$.  When $\lambda v^2\leq  R-v$, both types of equilibria may exist. From a semi-responsive equilibrium, the principal's expected payoff is again $W^{FT}=   2\pi \phi \max \{\sqrt{\lambda (R-v)}, \lambda v\}v -\pi \phi v+\alpha \pi [1+(1-\pi)\phi]>2 \pi \phi \lambda v^2-v+\alpha \pi=W^{FT}$. From an unresponsive  equilibrium, the principal's expected payoff is minimized when $e^*=\lambda v$, which is exactly equal to $W^{NT}$. 

 (2)
Now we compare the principal's expected payoffs $W^{ST}$ and $W^{FT}$. When a semi-responsive equilibrium is selected under \textit{Full-Transparency},  the principal receives a strictly higher political payoff than under \textit{Semi-Transparency} ($\alpha \pi [1+(1-\pi)\phi]> \alpha \pi[1+(1-\pi)\phi \lambda v ]$). This is because the principal can perfectly identify the type $(C,G)$ from a semi-responsive equilibrium under \textit{Full-Transparency}, while she can learn 
about the types only from a noisy signal (reform outcomes) under \textit{Semi-Transparency}. 

[I] If $\lambda v^2>R-v$, then a semi-responsive equilibrium exists and is unique under \textit{Full-Transparency}. In this case, the type $(C,G)$ agent reforms with implementation effort $\lambda v$, which is strictly lower than his effort under \textit{Semi-Transparency}. Thus, the principal receives a strictly higher policy payoff under  \textit{Semi-Transparency} than \textit{Full-Transparency}. She prefers  \textit{Full-Transparency} to \textit{Semi-Transparency} only if the political payoff more than offset the difference in policy payoffs ($\alpha$ is sufficiently large). If $\alpha$ is relatively small, the principal prefers \textit{Semi-Transparency}. 

[II] If  $\lambda v^2\leq R-v$, then both types of equilibrium may exist under \textit{Full Transparency}. When a semi-responsive equilibrium is selected, the type $(C,G)$ agent exerts higher implementation effort under  \textit{Full-Transparency} than \textit{Semi-Transparency} when both $\lambda v<1/3$ and $R\in (\underline{R}, \Bar{R})$ hold. In this situation, the principal receives a strictly higher payoff from \textit{Full-Transparency} than \textit{Semi-Transparency} when $\pi$ is large. This is because she receives both a strictly higher policy payoff and a strictly higher political payoff when the agent is likely congruent. When an unresponsive equilibrium is selected where all types choose to reform with implementation $e^*$, the principal receives a strictly lower political payoff from \textit{Full-Transparency} than \textit{Semi-Transparency}. She  receives a strictly higher policy payoff from \textit{Full-Transparency} than \textit{Semi-Transparency} if and only if (a) $\lambda v<1/3$ and $R\in (\underline{R}, \Bar{R})$, and $e^*$ is sufficiently close to $\sqrt{\lambda (R-v)}$. (b) a good reform  is very likely to occur ($\phi$ is sufficiently large). When all the condition hold, \textit{Full-Transparency} is preferable when the principal's policymaking motive dominates her political motive. That is, when $\alpha$ is sufficiently small.  In all other situations, the principal prefers \textit{Semi-Transparency} to \textit{Full-Transparency}.

As a sanity check, we examine whether Condition M is compatible with the assumptions in the benchmark model. By sending $v\rightarrow 0$, Condition M boils down to the inequality $R\in (0, \frac{4}{\lambda})$. This condition is guaranteed to hold by Assumptions in Section \ref{IA}, where $\lambda (v+\frac{R}{2})\leq 1$ reduces to $R< \frac{2}{\lambda}$ for $v\rightarrow 0$. 
\end{proof}

\subsection{Proof of Proposition \ref{Prop:AoN}}
\begin{proof}
A nonacountable official's optimal policy choice is to maintain the status quo, which brings the principal a policy payoff of $0$ and a political payoff of $\alpha \pi$. By allocating reform decisions to an accountable official with the information structure \textit{Semi-Transparency}, the principal receives a payoff of $W^{ST}= 2\pi  \phi \lambda v^2 +\phi\lambda Rv-\phi v+\alpha \pi[1+(1-\pi)\phi \lambda v]$, which is strictly higher than $\alpha\pi$ for $\lambda R \geq 1$.

As a sanity check, $\lambda R\geq 1$ is consistent with Assumptions in Section \ref{IA} for a nondegenerate set of parameters. By setting $v\rightarrow 0$, the assumptions simplify to $R\in [\frac{1}{\lambda}, \frac{2}{\lambda})$, which is compatible with a large set of parameters $(R,\lambda)$. 
\end{proof}

\subsection{Proof of Proposition \ref{Centralization}}
\begin{proof}
We suppose that $\underline{e}$ exceeds $\lambda v$. Otherwise, centralization brings the principal a payoff weakly lower than under \textit{Non-Transparency}. 

By centralizing, the principal receives a payoff of $ 2 \phi \underline{e}  v -v+   \alpha \pi$. By delegating to the agent along with the \textit{Semi-Transparency} information structure, the principal receives a payoff of $W^{ST}= 2\pi  \phi \lambda v^2 +\phi\lambda Rv-\phi v+\alpha \pi[1+(1-\pi)\phi \lambda v]$.  Thus, the principal should reclaim the decision making authority only if 1) the agent is unlikely to be congruent ($\pi$ is sufficiently close to $0$), 2) the reform is highly likely to be good (high $\phi$), 3) the level of enforceable implementation effort $\underline{e}$ is sufficiently close to 1, and 4) the political weight $\alpha$ is sufficiently close to 0. 
\end{proof}

\section{Model Robustness}\label{Rob}

\subsection{Weaker office-holding motives}\label{R1: office}
In Section \ref{IA}, we suppose that  $R \geq \max\{2\sqrt{\frac{v}{\lambda}},v\}$. The condition $R\geq v$ is commonly imposed to ensure that the agent has pandering incentives. Without this condition, the agent would likely select his favored policies without considering its implications for his career. This renders  the problem uninteresting. Conversely, the condition $R \geq 2\sqrt{\frac{v}{\lambda}}$ is less intuitive. We delve deeper into its implication in this subsection.

Suppose  $2\sqrt{\frac{v}{\lambda}}> R \geq v$. While the proofs for Propositions \ref{Prop:NT} and \ref{Prop:FT} remain intact, the agent does not necessarily select policies contingent on the state of the world under \textit{Semi-Transparency}, as per Proposition \ref{Prop:ST}. Thus, we conjecture an equilibrium of the following form:  
\begin{itemize}
    \item A type $(C,G)$ reforms with implementation effort $e^*_H$.
\item A type $(N,G)$ reforms with implementation effort $e^*_L$ with probability $\kappa\in (0,1]$, and maintains the status quo with probability $1-\kappa$. 
\item Types $(t, B)$ maintain the status quo.
\item The principal retains surely following a successful reform, and retains with probability $\sigma\in [0,1]$ after an unsuccessful reform. She replaces the agent after observing the status quo. 
\end{itemize}

The conditions for equilibrium are:
\begin{align}
\pi&=\frac{\pi (1-e^*_H)}{\pi  (1-e^*_H)+(1-\pi) (1-e^*_L)\kappa} \label{Ind}\\
  e^*_H &\in \max e(R+v)+(1-e)(R\sigma -v)-\frac{e^2}{\lambda }\label{optH}\\
  e^*_L &\in \max e R+(1-e)R\sigma-v-\frac{e^2}{\lambda }\label{optL}.\\
      R\cdot \sigma & <  v = e_L^* R+(1-e_L^*)R\sigma -\frac{{e^*}^2_L}{\lambda}  \label{Bt}.
\end{align}

Simplifying this system of equation, we have:
\begin{align*}
 1-e_H^*&=(1-e_L^*)\kappa\\
 e_H^*&= \lambda (v+\frac{1-\sigma}{2}R)\\
  e_L^*&= \lambda (\frac{1-\sigma}{2}R)\\
  v&=\lambda R^2(\frac{1-\sigma}{2})^2 + R\sigma.
\end{align*}
This leads to $\sigma=\frac{\lambda R/2-1+\sqrt{1-\lambda (R-v)}}{\lambda R/2}\in (0,1), \kappa=\frac{1-\lambda (v+\frac{1-\sigma}{2}R)}{1-\lambda (\frac{1-\sigma
}{2}R)}\in (0,1)$.

Continuing with our equilibrium characterization: Suppose the principal observes a successful reform, then she believes that it is more likely to come from type $(C,G)$ than $(N,G)$, by the Bayes rule. If instead she observes a failed reform, then Condition \ref{Ind} ensures that the principal's retention strategy is sequentially rational. Moreover, Conditions \ref{optH}-\ref{Bt} ensure that different types of the agent choose their optimal policies and implementation effort. Thus, no one has incentives to deviate from the conjectured equilibrium. 

It is useful to compare conditions \ref{Ind}-\ref{Bt} to the conditions that guarantee the equilibrium of Proposition \ref{Prop:ST}. In Proposition \ref{Prop:ST}, the principal retains whenever she observes a successful reform. Here, the principal retains surely whenever she observes a successful reform and probabilistically whenever she observes a failed reform. 
Essentially, a noncongruent agent's gamble for a successful reform is justified if the office rent is sufficiently large. If not (as suggested by $2\sqrt{\frac{v}{\lambda}}>R$), failed reforms should also sometimes be rewarded to incentivize the noncongruent agent to pursue reform, as justified by Condition \ref{Ind}. Else, the equilibrium structures in both Proposition \ref{Prop:ST} and this analysis are similar, indicating that the parameters of $R$ do not qualitatively alter the outcomes.


\subsection{Asymmetric policy stakes}\label{R2: asymmetric}
In Section 2.3 of the main text, we set parameter assumptions to ensure that considering the cost of implementation effort, the agent's optimal policy for all types is to maintain the status quo. We choose this specification primarily for clarity and simplicity in exposition. It does not significantly alter the core insights of our model.

To elaborate on this point, we suppose that the agent values maintaining the status quo at  $d\neq 0$. If $d<0$,  the congruent type may find reform preferable to the status quo despite implementation costs. Specifically, we can set $d$ such that $d<\lambda v^2-v<0$, which ensures that a type $(C,G)$ agent's optimal policy choice is to initiate a reform. Let's examine whether  this adjustment might influence the equilibrium across different information structures:
\begin{itemize}
    \item Under \textit{Non-Transparency}, the agent panders to reform and choose the implementation effort described in Proposition \ref{Prop:NT}, provided that $R-d\geq v$. Comparing this with Proposition \ref{Prop:NT}, the only difference is that the value of retention relative to maintaining the status quo shifts to $R-d$, as opposed to $R$ in the benchmark model.  
    \item Under \textit{Semi-Transparency}, a type $(\cdot, G)$ will reform and choose the implementation effort described in Proposition \ref{Prop:ST}, and a type $(\cdot, B)$ agent will choose the status quo, provided that $\lambda (R/2)^2-v\geq d$. Comparing this condition with  $\lambda (R/2)^2-v\geq 0$ in the benchmark model, there is no substantive difference. 
    \item Under \textit{Full-Transparency}, either a pooling equilibrium or a semi-separating equilibrium may exist. The least costly effort $e'_S$ that achieves type separation is determined by the equation $R-v-d=\frac{{e'}_S^2}{\lambda}$, or $e'_S=\sqrt{\lambda (R-v-d)}$, as opposed to $e_S=\sqrt{\lambda (R-v)}$ in the benchmark model. To determine whether a pooling equilibrium or a semi-separating equilibrium may arise, one will compare whether $\lambda v^2\geq R-v-d$ is true, as  opposed to $\lambda v^2\geq R-v$ in the benchmark model. Again, there is no substantive difference. 
\end{itemize}

Thus, we conclude that whether the value $d$ is set to zero (benchmark) or not (extension) does not  substantively affect the agent's equilibrium behavior. A $d$ smaller than zero renders the status quo less appealing to the principal, making her more receptive to change. Nonetheless, this does not substantively affect her optimal choice of the information structure.

\subsection{More data on implementation}\label{R3:data}
In the main text, the ``moral hazard'' in reform policymaking is encapsulated using the implementation effort $e$. Now we relax this assumption.

Consider a more general environment in which the reform implementation involves a vector of necessary inputs $(e_1,e_2,...e_n)$.  Each  of these inputs defines the success probability $h\in [0,1]$ of a {\it good} reform according to the function $h=h(e_1,e_2,...e_n)$. Consistent with our previous assumptions, a \textit{bad} reform always fails. Let $C(e_1,e_2,...e_n)$ be the cost function. Then, conditional on a reform decision, a congruent agent without retention incentive will solve the following optimization problem:
\begin{align*}
    \max_{e_1,e_2,...e_n} (2\mu h(e_1,e_2,...e_n)-1)v-C(e_1,e_2,...e_n),
\end{align*}
where $\mu$ denotes whether the reform is good ($\mu=1$) or bad ($\mu=0$). In scenarios where retention is contingent on reform success, an agent will solve the following optimization problem:
\begin{align*}
  & \text{Congruent}\qquad  \max_{e_1,e_2,...e_n} \mu h(e_1,e_2,...e_n)(2v+R)-v-C(e_1,e_2,...e_n)\\
   & \text{Nonongruent}\qquad \max_{e_1,e_2,...e_n} \mu h(e_1,e_2,...e_n)R-v-C(e_1,e_2,...e_n)
\end{align*}
The solutions to the above optimization programs follow the standard textbook procedure (up to regularity conditions): suppose an agent wants to target a level of success probability $\Bar{h}$. He shall solve $\min_{e_1,...e_n} C(e_1,e_2,...e_n)\quad s.t.\; h(e_1,e_2,...e_n)\geq \bar{h}$. This gives us an induced cost function for $h$, which we denote by $C(h)$ and effectively translates our multivariate moral hazard issue into a univariate representation. Now we relate this representation to the equilibrium analysis under different information structures.

When the implementation details remain hidden from the principal (under \textit{Non-Transparency} or \textit{Semi-Transparency}), whether the complexity of the dimension (be it one or more) becomes inconsequential. An agent will optimally choose the input vector $(e_1,e_2,...e_n)$ to target his desired success probability $h$. The situation differs slightly under \textit{Full-Transparency}, as the agent can signal types by choosing the input vector $(e_1,e_2,...e_n)$. Yet, using an argument analogous to that in our main text, two equilibria will prevail depending on the parameters: 
\begin{enumerate}
    \item When the office rent $R$ is not sufficiently large, a type $(C,G)$ agent will reform while other types choose the status quo.
    \item 
 When the office rent $R$ is sufficiently large, either (1) a type $(C,G)$ agent will reform while other types choose the status quo, or (2) all types pool on reform with a vector of input $(e_1,e_2,...e_n)$ that guarantees a sufficiently high likelihood of successful reform. 
\end{enumerate}

Given this, we conclude that the baseline model entails no loss of generality.

\subsection{Alternative information structure}\label{R4:Alternative}
In our baseline model, we exclude an information structure where the principal observes the policy choice and implementation efforts, but not the consequences. This structure allows the principal to glean more information about policymaking than \textit{Non-Transparency}, less than \textit{Full-Transparency}, but not directly comparable to \textit{Semi-Transparency}. In this subsection, our goal is to rank this information structure in terms of the principal's welfare.

As we have explicated in Proposition \ref{Prop:FT}, when the principal has observed the policy choice, the implementation details, and policy outcomes, she will discard the information from policy outcomes in the retention decision. 
 Put simply, it is the observable actions rather than their induced policy outcomes are crucial for the principal's retention decisions.  Therefore, once we exclude the policy outcomes from what the principal observes,  the agent faces exactly the same policymaking and career incentives as under \textit{Full-Transparency}.  The conditions supporting the conclusions of Proposition \ref{Prop:FT} continue to support the same equilibrium policymaking behaviors when the policy outcomes
 are concealed from the principal. Thus, studying this information structure would be repetitive.

Technically, this ``outcome irrelevance'' result comes from the fact that agents' actions are sufficient statistics for their types: once the agents' actions are observed, the resulting policy outcomes do not provide any additional information about the agent's private type. Thus, under \textit{Full-Transparency}, the principal will not condition her retention decision on the policy outcomes. Neither can the principal condition her retention decision on the policy outcomes under \textit{Non-Transparency}, where outcomes are unobserved. This implies that only under \textit{Semi-Transparency} can the principal implement a ``performance-based'' retention rule that  links the agent's policy and political payoffs.

\subsection{Modeling congruence}\label{R5:congruence}
We model (non)congruence along the line of \cite{fox2007government}. This naturally leads to the question: would alternative notions of congruence yield similar or different theoretical predictions? We explore this possibility here with a simplified model.

Our main job is to illustrate that the agent's political and policy payoffs can be linked only under \textit{Semi-Transparency}. After that, we explore the implications of institutional design on welfare  under this alternative notion of congruence. Consistent with our baseline model,  the agent observes the state of a reform $\omega$ prior to taking action. Good reform decisions and good implementation are complementary to a successful reform. That is, a bad reform always fails; a good reform succeeds with probability equal to the implementation effort. To simplify matters, the effort is chosen from the binary set $\{\underline{e}, 1\}$ with the associated costs $c(\underline{e})=0$ and $ c(1)=c>0$.

Per \cite{maskin2004politician}, we suppose that different types of the agent have opposing state-dependent policy preferences.  Just as the principal, the congruent type receives $v>0$ from a successful reform, $0$ from the status quo, and $-v$ from a failed reform; the noncongruent type receives  $-v$ from a successful reform, $0$ from the status quo, and $v$ from a failed reform. We impose $\underline{e}>\frac{1}{2}$ to ensure that the noncongruent type averts a reform when $\omega=G$.

The timeline is the same as the baseline model. 
\begin{enumerate}
    \item Nature picks the random variables $(\omega,t)$ according to the prior distributions. 
    \item The agent observes $\omega,t$, and chooses $x\in\{r,q\}$. Conditional on $x=r$, he chooses implementation effort $e\in\{\underline{e},1\}$.
    \item  The policy consequence $y$ is realized. 
    \item The principal decides whether to retain the agent conditional on her available information. 
    \item All players' payoffs are realized.
\end{enumerate}

We impose the following assumptions. First, $\phi>\frac{1}{2}$. Similar with \cite{maskin2004politician}, reform is the ex ante popular action. Second,  $2(1-\underline{e})v<c<\min \{v, (1-\underline{e})(2v+R)\}$. It ensures that the congruent type reforms diligently  only when political and policy payoffs are linked. Third, $v+c>R>v$. It guarantees that the noncongruent type may pander to a reform but will not expend costly efforts in implementation. Since $\underline{e}>1/2$, for any given parameters $(R,v)$ with $R>v$, a continuum values of $c$ is compatible with the assumptions.

We characterize the agent's equilibrium behaviors under three information structures: \textit{Non-Transparency}, (\textit{NT}), \textit{Semi-Transparency} (\textit{ST}), and \textit{Full-Transparency} (\textit{FT}):
\begin{Proposition}
\begin{enumerate}
    \item (\textit{NT}) and  (\textit{FT})  beget inaction: every type of the agent chooses the status quo regardless of $\omega$. 
    \item (\textit{ST}) begets motivation. Specifically,
    \begin{itemize}
        \item If $R<2v$, the type $(C,G)$ agent reforms with effort $1$. The type $(C,B)$ maintains the status quo. The type $(N,G)$ agent reforms with effort $\underline{e}$ with probability $\kappa^*=2-\frac{1}{\phi}$ and maintains the status quo with probability $1-\kappa^*$. The type $(N,B)$ agent reforms with effort $e=\underline{e}$. The principal retains with probability $1$ after observing a successful reform,  with probability $0$ after observing a failed reform, and  with probability $\sigma^*=\underline{e}-(2\underline{e}-1)\frac{v}{R}$ after observing the status quo. 
        \item If $R\geq 2v$, the type $(C,G)$ agent reforms with effort $1$. The type $(C,B)$ maintains the status quo. The type $(N,G)$ agent reforms with effort $\underline{e}$. The type $(N,B)$ agent maintains the status quo. The principal retains with probability $1$ after observing a successful reform,  with probability $0$ after observing a failed reform, and  with probability $\sigma^*=\frac{v}{R}$ after observing the status quo. 
    \end{itemize}
\end{enumerate}
\end{Proposition}
\begin{proof}
(1a). Under (\textit{NT}),  the principal's posterior belief is that $P(t=C|x=q)=\pi$ by the Bayes rule. Off the equilibrium path, the divinity criterion assigns belief $P(t=C|x=r)=0$. This is because the type $(N,B)$ agent may 
profitably deviate from the status quo to pursue a failed reform; further more, this type has the most to gain
from such a deviation. In comparison, the type $(C,B)$ cannot successfully reform, while $(C,G)$ must expend costly effort to achieve the desired successful reform. Thus, it is sequentially rational for the principal to retain after observing the status quo and replace after observing reforms. Given this retention rule, the agent's optimal response is to choose the status quo regardless of types. By deviation, one would obtain a payoff of at most $v$, which is lower than the equilibrium payoff $R$.

(1b). Under (\textit{FT}),  the principal's posterior belief is that $P(t=C|x=q)=\pi$ by the Bayes rule. Off the equilibrium path, the divinity criterion assigns belief $P(t=C|x=r, \cdot)=0$. As we have explicated in (1a), this is because a type $(N,B)$ agent has the most to gain by deviating from the status quo (to pursue a failed reform). Along the path, the principal retains after observing the status quo and replaces after observing reforms. Given this retention rule, the agent's optimal response is to choose the status quo regardless of types. By deviation, one would obtain a payoff of at most $v$, which is lower than the equilibrium payoff $R$.

(2a). [$R<2v$] Under (\textit{ST}), the principal's posterior beliefs are $P(t=C|r,S)=\frac{\pi}{\pi+(1-\pi)\underline{e}\kappa^*}>\pi$ and $P(t=C|r,F)=0$.  Furthermore, $P(t=C|x=q)=\frac{\pi (1-\phi)}{\pi(1-\phi)+(1-\pi)\phi(1-\kappa^*)}=\pi$. The principal retains surely if she observes a successful reform and replaces surely if she observes a failed reform; she retains with probability $\sigma^*\in [0,1]$ after observing the status quo. Under this incentive scheme, [1] the type $(C,G)$ agent cannot benefit from deviating to the status quo. If he chooses to do so, his payoff will change from $R+v-c$ to $R\sigma^*$, which is equal to $(R-2v)\underline{e}+v<v$. He does not want to reduce effort to $\underline{e}$ either, as the saved implementation cost $c$ does not offset the forgone policy and political payoff $(1-\underline{e})(R+2v)$. [2] The type $(C,B)$ agent does not want to deviate to the reform, for his reform cannot succeed, and his reelection probability will be lower.
[3] The type $(N,G)$ agent randomizes in equilibrium. He will not deviate to increasing his implementation effort to $1$, as this is strictly dominated.  Randomization requires this type to be indifferent between $(r, \underline{e})$ and $q$. This means that in equilibrium, the probability that the principal retains after observing the status quo, $\sigma^*$, must satisfy $R\underline{e}-(2\underline{e}-1)v=R\sigma^*$. This pins down the equilibrium $\sigma^*=\underline{e}-(2\underline{e}-1)\frac{v}{R}$, which is strictly smaller than $1$ under the assumption $\frac{v}{R}<1$. [4] The type $(N,B)$ agent does not want to deviate to choosing the status quo. His equilibrium payoff is $v$, which is larger than $R\underline{e}-(2\underline{e}-1)v$ if and only if $R<2v$. Thus, if the type $(N,G)$ randomizes, the type $(N,B)$ will choose to reform with implementation effort $\underline{e}$ surely.

(2b). [$R\geq 2v$] Under (\textit{ST}), the principal's posterior beliefs are $P(t=C|r,S)=\frac{\pi}{\pi+(1-\pi)\underline{e}}>\pi$ and $P(t=C|r,F)=0$.  Furthermore, $P(t=C|x=q)=\pi$. The principal retains surely if she observes a successful reform and replaces surely if she observes a failed reform; she retains with probability $\sigma^*\in [0,1]$ after observing the status quo. Under this incentive scheme, [1] the type $(C,G)$ agent cannot benefit from deviating to the status quo, for his payoff will decrease from $R+v-c$ to $v$. He does not want to reduce effort to $\underline{e}$ either, as the saved implementation cost $c$ does not offset the forgone policy and political payoff $(1-\underline{e})(R+2v)$. [2] The type $(C,B)$ agent does not want to deviate to the reform, for his reform cannot succeed, and his reelection probability will be lower.
[3] The type $(N,B)$ agent does not benefit from deviating to reform with effort $\underline{e}$. This is because his equilibrium payoff is $R\sigma^*$, while his payoff from this deviation is $v$. Thus, the type $(N,B)$ agent is indifferent between two actions. [4] The type $(N,G)$ agent does not benefit from deviating to increasing his implementation effort to $1$, as this is strictly dominated.  He does not benefit from deviating to the status quo.  This is because his equilibrium payoff is $R\underline{e}-(2\underline{e}-1)v$,
which is greater than  $v$ if $R\geq 2v$. 
\end{proof}

From the above analysis, the principal's payoff is $0$ under both (\textit{NT}) and (\textit{FT}). Her payoff under  (\textit{ST}) is $[\phi \pi  +(1-\pi)(4\phi\underline{e}-2\underline{e}-\phi)]v$ if $R<2v$ and $\phi[\pi+(1-\pi)(2\underline{e}-1)]v$ if $R\geq 2v$.  In the former case,  (\textit{ST}) is the principal-optimal institution design whenever $\phi\pi+ (1-\pi)(4\phi\underline{e}-2\underline{e}-\phi)\geq 0$. In the latter, (\textit{ST}) is always optimal. 

\end{appendices}
\end{document}